\documentclass[5pt]{aastex}  

\newcommand{\bzcat}{Roma-BZCAT}

\newcommand{\chn}{{\it Chandra}}

\newcommand{\fer}{{\it Fermi}}

\newcommand{\swf}{{\it Swift}}

\newcommand{\xmm}{{\it XMM-Newton}}
\newcommand{\wse}{{\it WISE}}
\usepackage{graphicx}
\usepackage{longtable}
\usepackage{mdwlist}
\usepackage[switch, modulo]{lineno}
\usepackage{amssymb}
\usepackage{lscape}
\usepackage{tablefootnote}

\usepackage{natbib}
\usepackage{layout}
\bibpunct{(}{)}{;}{a}{}{,} % to follow the A&A style
\usepackage{multirow}
\usepackage{upgreek}
\usepackage[colorlinks=false, pdfstartview=FitV, linkcolor=blue,citecolor=blue, urlcolor=blue]{hyperref}
\usepackage{wasysym}
\usepackage{txfonts}
\usepackage{gensymb}
\usepackage{morefloats}
\usepackage{threeparttable}
\usepackage{graphicx}
\usepackage{lscape}

\begin{document} 
\title{Optical spectroscopic observations of $\gamma$-ray blazar candidates III. The 2013/2014 campaign in the Southern Hemisphere\thanks{Based on observations obtained at the Southern Astrophysical Research (SOAR) telescope, which is a joint project of the Minist\'{e}rio da Ci\^{e}ncia, Tecnologia, e Inova\c{c}\~{a}o (MCTI) da Rep\'{u}blica Federativa do Brasil, the U.S. National Optical Astronomy Observatory (NOAO), the University of North Carolina at Chapel Hill (UNC), and Michigan State University (MSU).}}

 \author{M. Landoni}
 \affil{INAF-Osservatorio Astronomico di Brera, Via Emilio Bianchi 46, I-23807 Merate, Italy}
 \affil{INFN - Istituto Nazionale di Fisica Nucleare}
 \affil{Harvard - Smithsonian Center of Astrophysics, 60 Garden Street, Cambridge, MA 02138, USA}
\affil{marco.landoni@brera.inaf.it}
\author{F. Massaro}
\affil{Yale Center for Astronomy and Astrophysics, Physics Department, Yale University, PO Box 208120, New Haven, CT 06520-8120, USA}
\affil{Dipartimento di Fisica, Universit\`a degli Studi di Torino, via Pietro Giuria 1, I-10125 Torino, Italy}
\author{A. Paggi}
 \affil{Harvard - Smithsonian Center of Astrophysics, 60 Garden Street, Cambridge, MA 02138, USA}
\author{R. D'Abrusco}
 \affil{Harvard - Smithsonian Center of Astrophysics, 60 Garden Street, Cambridge, MA 02138, USA}
\author{ D. Milisavljevic}
 \affil{Harvard - Smithsonian Center of Astrophysics, 60 Garden Street, Cambridge, MA 02138, USA}
 \author{N. Masetti}
\affil{INAF - Istituto di Astrofisica Spaziale e Fisica Cosmica di Bologna, via Gobetti 101, 40129, Bologna, Italy}
\author{H. A. Smith}
 \affil{Harvard - Smithsonian Center of Astrophysics, 60 Garden Street, Cambridge, MA 02138, USA}
 \author{G. Tosti}
 \affil{Dipartimento di Fisica, Universit\`a degli Studi di Perugia, 06123 Perugia, Italy}
 \author{L. Chomiuk}
 \affil{Department of Physics and Astronomy, Michigan State University, East Lansing, MI 48824, US}
 \author{J. Strader}
  \affil{Department of Physics and Astronomy, Michigan State University, East Lansing, MI 48824, US}
\author{C. C. Cheung}
\affil{Space Science Division, Naval Research Laboratory, Washington, DC 20375, USA}

%\date{Received September 1, 2014; accepted  ... }

 \begin{abstract}
{We report the results of our exploratory program carried out with the Southern Astrophysical Research (SOAR) telescope aimed at associating counterparts and establishing the nature of the Fermi Unidentified $\gamma$-ray Sources (UGS). We selected the optical counterparts of 6 UGSs from the \fer\ catalog on the basis of our recently discovered tight connection between infrared and $\gamma$-ray emission found for the $\gamma$-ray blazars detected by the Wide-Field Infrared Survey Explorer (WISE) in its the all-sky survey. We perform for the first time a spectroscopic study of the low-energy counterparts of \fer\ UGS, in the optical band, confirming the blazar-like nature for the whole sample.
We also present new spectroscopic observations of 6 Active Galaxies of Uncertain type associated with \fer\ sources  (AGUs) that appear to be BL Lac objects.
Finally, we report the spectra collected for 6 known $\gamma$-ray blazars belonging to the Roma BZCAT \citep{massaro09} that were obtained to establish their nature 
or better estimate their redshifts. Two interesting cases of high redshift and extremely luminous BL Lac objects ($z \geq 1.18$ and $z \geq 1.02$, based on the detection of Mg II intervening systems) are also discussed.}
\end{abstract}

\keywords{galaxies: active - galaxies: BL Lacertae objects -  radiation mechanisms: non-thermal}
%\authorrunning{M. Landoni et al}
%\maketitle

\section{Introduction}
{According to the Second \fer-LAT Source Catalog \citep[2FGL;][]{nolan12} 
and subsequent analyses performed on the Unidentified $\gamma$-ray Sources \citep[UGSs; e.g.,][]{paper4,acero13}
a significant fraction of the $\gamma$-ray sky (about 30\%) is yet unknown \citep{ugs2}.
For this reason,
we developed two association procedures \citep[e.g.,][]{ugs1,paper4} able to associate counterparts of \fer\ UGSs. These methods are
based on correlation between infrared (detected in the Wide-Field Infrared Survey Explorer \citep[\wse;][]{wright10}) and $\gamma$-ray emission \citep[e.g.,][]{paper1,paper2} of blazars which are the largest known population of $\gamma$-ray sources.
}

{Blazars are active galactic nuclei (AGNs) dominated by non-thermal radiation over the 
entire electromagnetic spectrum, presenting very peculiar observational properties with respect to other AGN classes. In particular they show
very rapid variability at all frequencies, high polarization, superluminal motion, and very high luminosities \citep[e.g.,][]{urry95}.
Their emission is interpreted as arising from ultra relativistic particles accelerated in a relativistic jet closely aligned 
to our line of sight \citep{blandford78}.
According to the literature, there are two classes of blazars objects. Briefly, the blazar subclass of BL Lac are characterised by optical spectrum where observed features have
rest frame equivalent width EW $\leq$ 5 $\AA$ \citep[e.g.,][]{stickel91,stoke91,laurent99} while the Flat Spectrum Radio Quasars (FSRQ) shows the typical quasar-like optical spectra characterized by strong and broad emission lines 
and higher radio polarization.}
%We note that in the \bzcat\ there are also several BZBs indicated as BL Lac candidates for which 
%no optical spectra were found in literature to guarantee a correct classification as well as sources 
%classified as blazars of uncertain type (i.e., BZUs) namely objects with  
%peculiar characteristics but showing blazar activity 
%(e.g., occasional presence/absence of broad spectral lines or features, transition objects between a
%radio galaxy and a BL Lac, galaxies hosting a low luminosity blazar nucleus, etc.).
%\\
\\
Here we report the results of our recent spectroscopic campaign in the Southern hemisphere
using the Southern Astrophysical Research (SOAR, see Clemens et al 2014) telescope. In this campaign we obtained spectra for optical counterparts of Fermi UGSs selected adopting the procedure previously  described in D'Abrusco et al 2013 and references therein. 
Preliminary results for our exploratory program in the Northern hemisphere 
obtained with the Telescopio Nazionale Galileo (TNG), 
the Multiple Mirror Telescope (MMT) and the Observatorio Astron\'omico Nacional (OAN) in San Pedro M\'artir (M\'exico) 
observatory have been already presented in Paggi et. al. 2014. Although additional multifrequency campaigns to investigate the origin of the UGSs are already ongoing, e.g., at the radio frequencies \citep[e.g.,][]{kovalev09,petrov13,ugs3,ugs6} or in the X-rays as the \swf\ X-ray survey of the UGSs\footnote{\underline{http://www.swift.psu.edu/unassociated/}} \citep[e.g.,][]{mirabal09,paggi13,takeuchi13,stroh13}, the optical spectroscopic campaigns (as the one presented in this serie of papers) are a crucial step, in particular for blazars, 
to disentangle and confirm the true nature of the low-energy counterparts selected  with different methods \citep[e.g.,][]{masetti13,shaw13a,shaw13b,paggi14,sdss, land0048, land13, land14}. 
{In this paper we also present spectroscopy for six Active Galaxy of Uncertain Type (AGU) in the \fer\ catalog and for six other known blazars in order to confirm their nature.
}

The paper is organized as follows: in Section~\ref{sec:obs} we discuss the data reduction procedures
adopted to analyze the SOAR observations acquired.
Then in Section~\ref{sec:results} we describe the results of our analysis providing details on each source.
Finally, the summary and conclusions are given in Section~\ref{sec:conclusions}.
We use cgs units unless stated otherwise and the following cosmological parameters $H_{0} = 70$ km s$^{-1}$    Mpc$^{-1}$, $\Omega_{m} = 0.27$, $\Omega_{\Lambda} = 0.73$.

\section{Observations and Data Reduction}
\label{sec:obs}

\begin{table*}
\scriptsize
\caption{Selected sample and observation log}
\begin{center}
%\resizebox{\textwidth}{!}{
\begin{tabular}{|lllllll||l|}
\hline
Name & \wse\ & R.A. & Dec. & Obs.\,Date & Exp. & notes & Class\\
           & name & (J2000) & (J2000) & (yy-mm-dd) & (min) &  &\\ 
\hline
& & & {UGS sample} & & && \\
\hline
2FGL J0116.6-6153 & J011619.62-615343.4 & 01:16:19.62 & -61:53:43.5 & 13-09-16 & 30 & S,w,U,M & BLL\\
2FGL J0133.4-4408 & J013306.37-441421.4 & 01:33:06.37 & -44:14:21.4 & 13-09-16 & 40 & S,w,U & BLL\\ 
2FGL J0143.6-5844 & J014347.41-584551.4 & 01:43:47.42 & -58:45:51.4 & 14-01-10 & 20 & S,w,M,U,u,x&BLL \\ 
2FGL J0316.1-6434 & J031614.34-643731.4 & 03:16:14.34 & -64:37:31.5 & 14-01-10 & 20 & S,w,M,U,u,x& BLL\\ 
2FGL J0416.0-4355 & J041605.82-435514.6 & 04:16:05.83 & -43:55:14.7 &13-09-16 & 40 & S,w,M,U& QSO\\ 
2FGL J0555.9-4348 & J055618.74-435146.0 & 05:56:18.74 & -43:51:46.0 & 13-09-16 & 40 & S,w,M,U& BLL\\ 
2FGL J2257.9-3646 & J225815.00-364434.3 & 22:58:14.65 &-36:44:38.0 & 13-11-28 & 20 & S,N,rf,w,u,x  & BLL \\
\hline
& & & {AGU sample} & & && \\
\hline
2FGL J0157.2-5259$^*$ & J015658.00-530200.0 & 01:56:57.75 & -53:01:57.5 & 13-08-03 & 20 & S,w,M,6,U,g,X,u,x  &BLL\\ %1FGL
2FGL J0335.3-4501 & J033513.88-445943.8 & 03:35:13.90 & -44:59:40.0 & 13-09-16 & 10 & S,w,U,X,g,6,u,x,U,X & BLL\\ 
2FGL J0424.3-5332 & J042504.27-533158.2 & 04:25:04.26 & -53:31:58.3 & 13-09-15 & 20 & Pm,S,A,c,w,M,U,X &BLL\\ 
2FGL J0537.7-5716$^*$ & J053748.96-571830.1 & 05:37:48.96 & -57:18:30.2 & 13-09-15 & 30 & S,w,M,U,X &BLL\\ %1FGL
2FGL J0604.2-4817$^*$& J060408.61-481725.1 & 06:04:08.62 & -48:17:23.6 & 13-09-16 & 20 & S,w,M,U,X,6 &BLL\\ %1FGL ??
2FGL J1103.9-5356 & J110352.22-535700.7 & 11:03:52.32 & -53:57:00.79 & 14-01-11 & 120 & Pm,Pk,A,M,w & BLL \\
\hline
& & & {$\gamma$-rays blazars} & & && \\
\hline
BZB J0158-3932 & J015838.10-393203.8 & 01:58:38.09 & -39:32:03.8 & 13-09-15 & 20 &--- &BLL \\ 
BZB J0237-3603 & J023734.04-360328.4 & 02:37:34.04 & -36:03:28.4 & 13-09-16 & 10 &--- &BLL\\ 
BZB J0238-3116 & J023832.48-311658.0 & 02:38:32.48 & -31:16:57.9 & 13-09-16 & 10 &--- &BLL\\ 
BZB J0334-4008 & J033413.65-400825.4 & 03:34:13.65 & -40:08:25.5 & 13-09-16 & 10 &--- &BLL\\ 
BZB J0428-3756 & J042840.41-375619.3 & 04:28:40.42 & -37:56:19.6 & 13-09-15 & 30 &---& BLL\\ 
BZB J1443-3908 & J144357.20-390840.0 & 14:43:57.21 & -39:08:39.9 & 13-09-15 & 5 &--- &BLL\\ 
\hline
%\hline
\end{tabular}
%}
\end{center}

- The asterisk ($^*$) close to the source name marks sources that are also associated in the 1FGL catalog \citep{abdo10}.\\
- Catalog/survey symbols in notes: PMN \citep[][- Pm]{wright94}, SUMSS \citep[][- S]{mauch03}, AT20G \citep[][- A]{murphy10},
CRATES \citep[][- c]{healey07}, WISE \citep[][- w]{wright10,cutri12}, 2MASS \citep[][- M]{skrutskie06}, 
USNO-B1 \citep[][- U]{monet03}, GALEX (- g), RBSC and RFSC \citep[][- X]{voges99,voges00}, Deep Swift X-Ray Telescope Point Source Catalog \citep[1SXPS;][- x]{evans14} and \swf\ X-ray survey for all the \fer\ 
UGSs$^9$ \citep{stroh13,paggi13,takeuchi13},Six-degree-Field Galaxy Survey \citep[6dFGS;][- 6]{jones04,jones09}
\label{tab:log}
\end{table*}
~

\begin{figure*}[htbp]
\centering
   \resizebox{18.5cm}{!}{\includegraphics{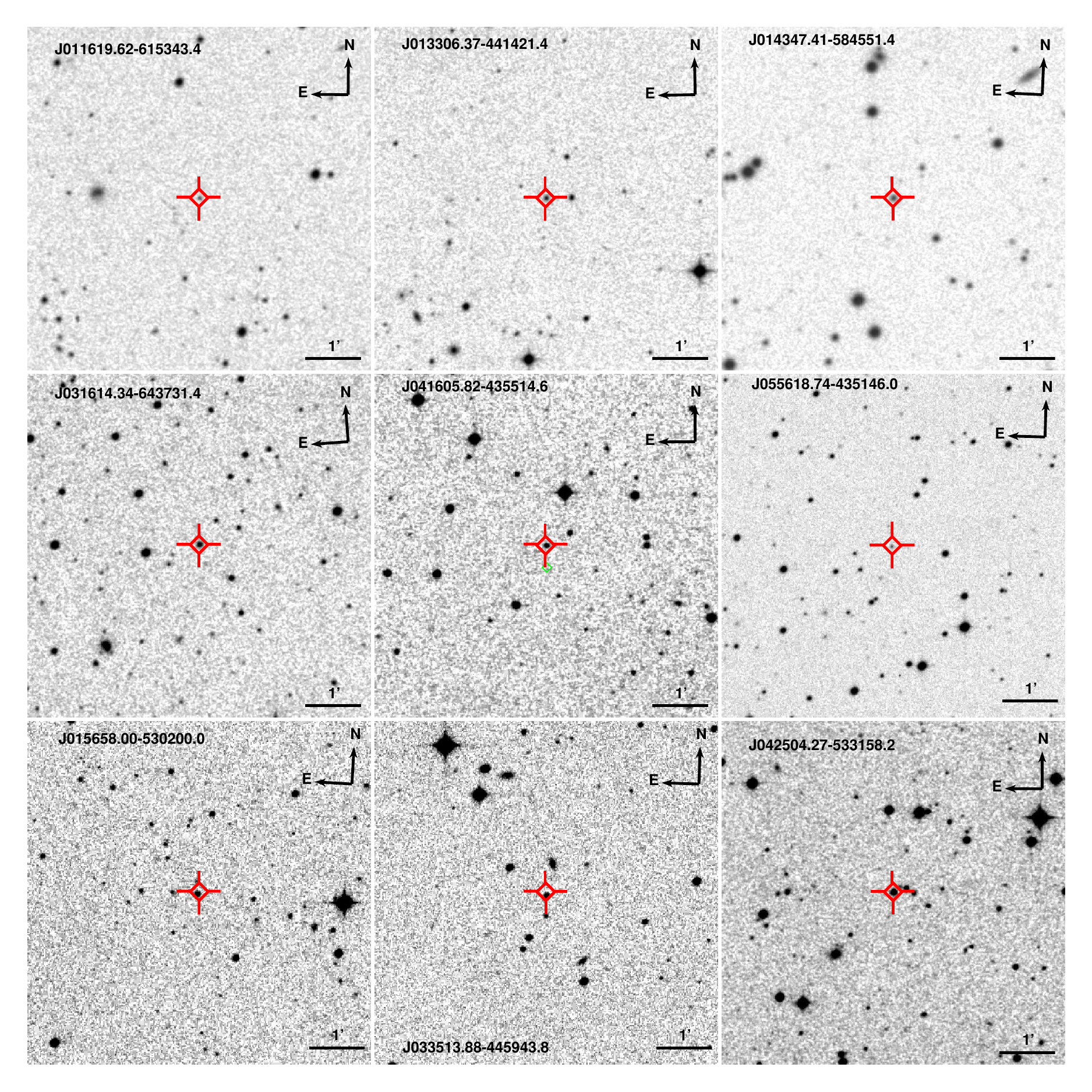}}
     \caption{Optical images of the fields of the WISE sources selected in this paper for optical spectroscopic follow-up (see Table 1). The object name, image
scale, and orientation are indicated in each panel. The proposed optical counterparts are indicated with red marks (WISE coordinates) and the fields are extracted from the DSS-II-Red
survey. A colour version of this figure is available on the online edition of the journal.}
     \label{fig:fc1}
\end{figure*}

\begin{center}
\begin{figure*}
   \includegraphics[width=18.5cm]{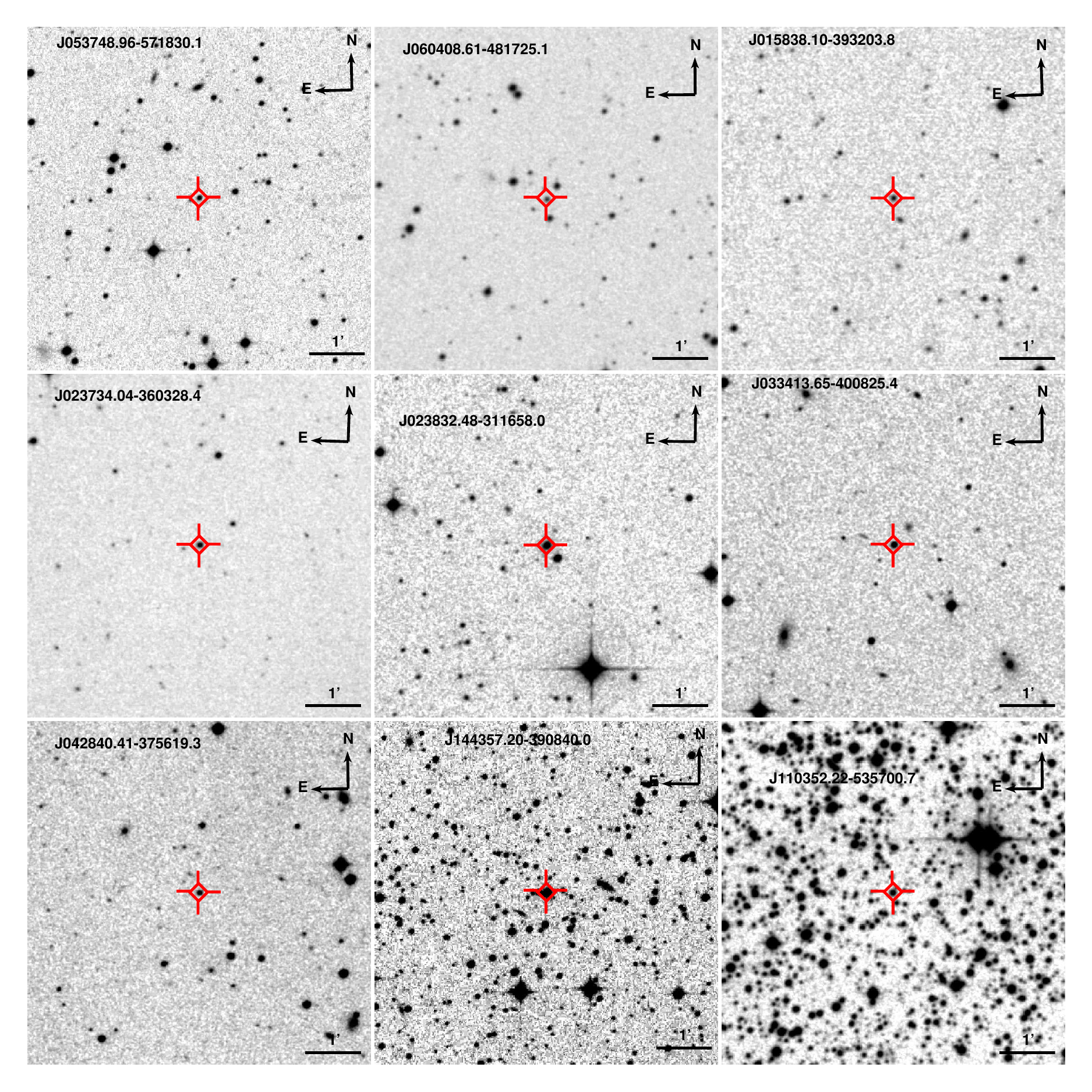}
     \caption{Same legend of Figure \ref{fig:fc1}}
     \label{fig:fc2}
\end{figure*}
\end{center}

\begin{center}
\begin{figure*}
   \includegraphics[width=5cm]{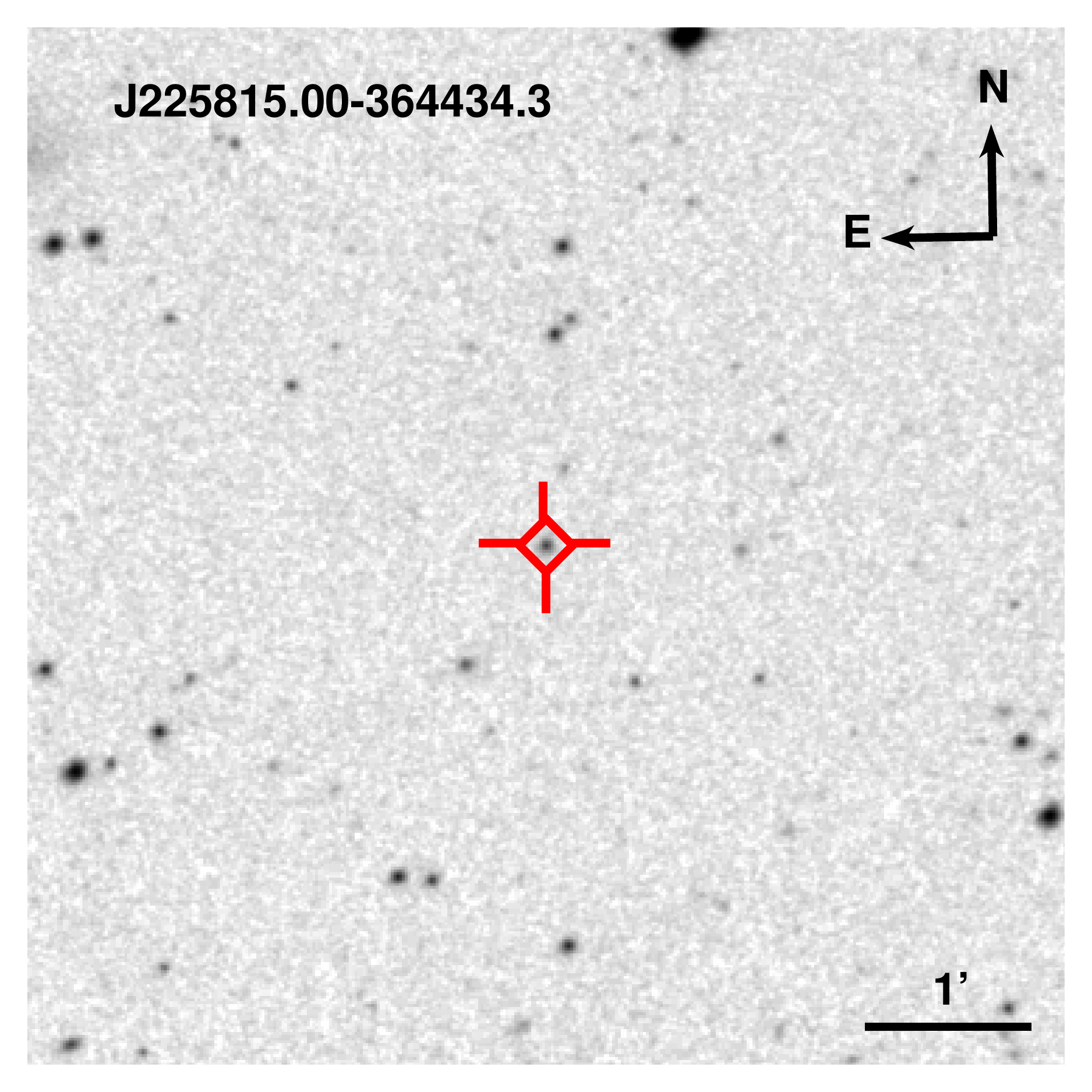}
     \caption{Same legend of Figure \ref{fig:fc1}}
     \label{fig:fc3}
\end{figure*}
\end{center}

\begin{center}
\begin{figure*}
   \includegraphics[width=12cm]{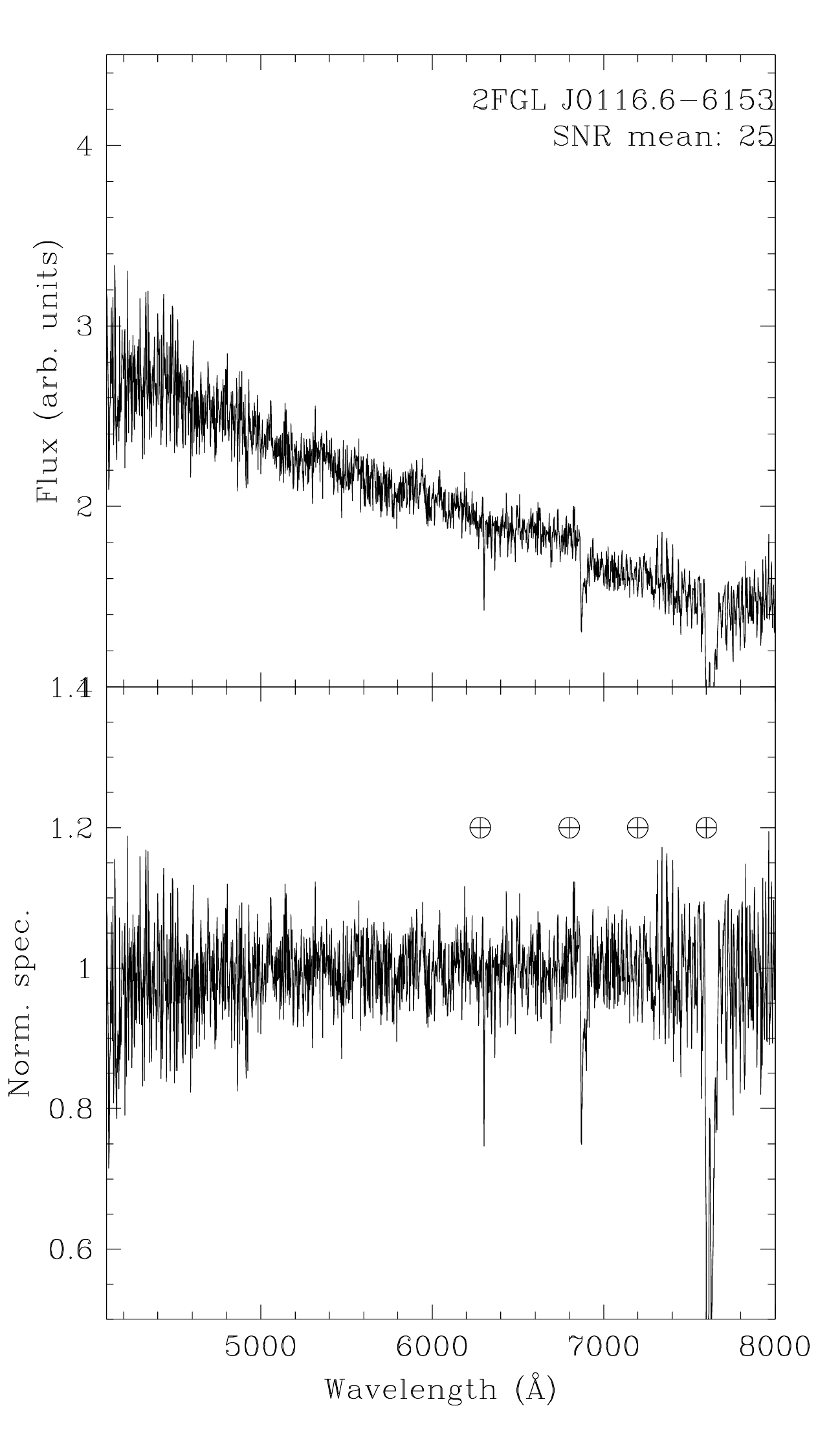}
     \caption{2FGL J0116.6-6153 optical spectrum obtained at SOAR with High Throughput Goodman Spectrograph. Upper panel: Flux calibrated (relative units) spectrum of the source. Lower panel: Normalised spectrum (see text for details).}
     \label{fig:2fglj0115}
\end{figure*}
\end{center}

\begin{center}
\begin{figure*}
   \includegraphics[width=12cm]{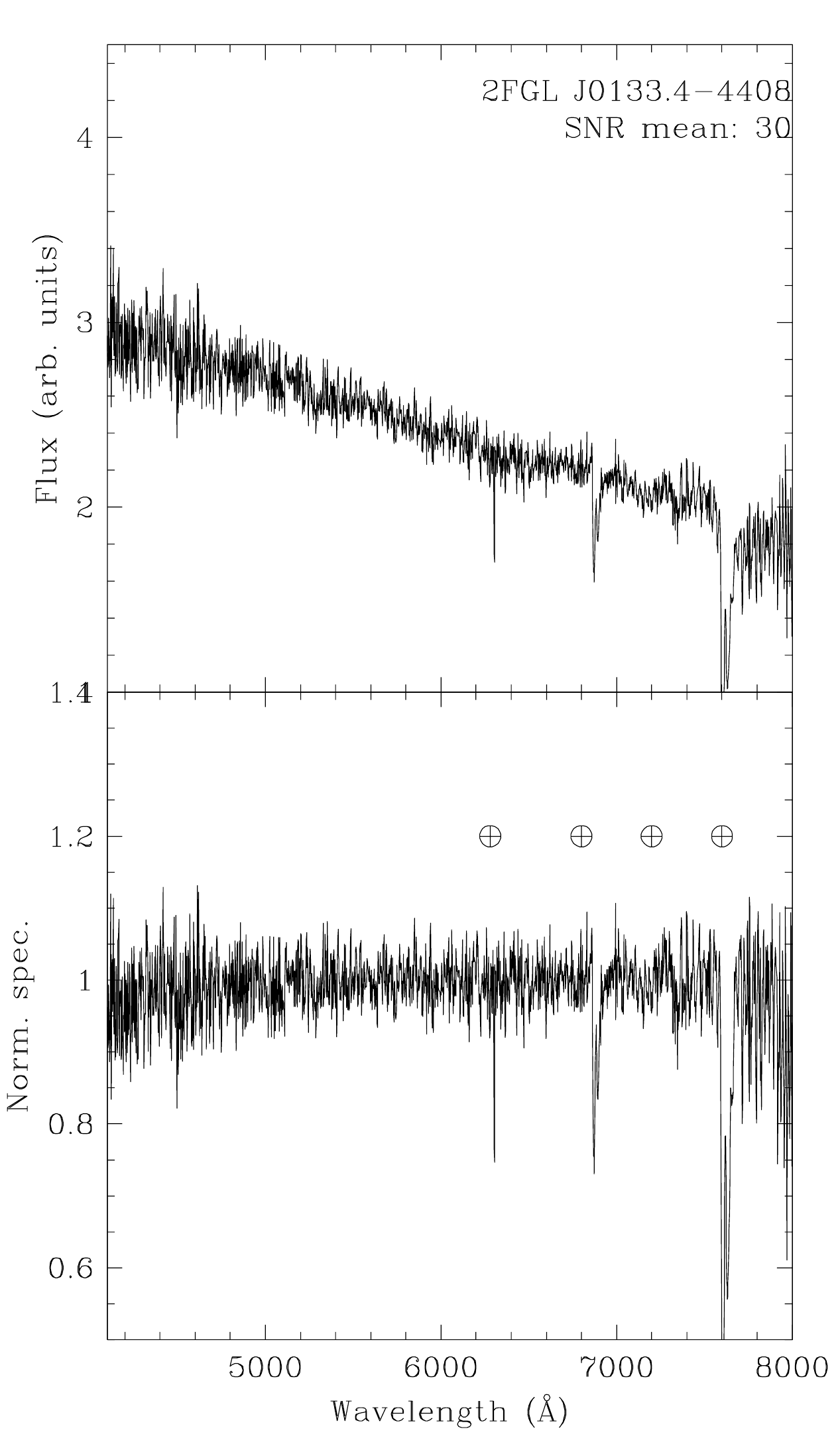}
     \caption{As in Figure \ref{fig:2fglj0115} but for 2FGL J0133.4-4408 }
     \label{fig:2fglj0133}
\end{figure*}
\end{center}

\begin{center}
\begin{figure*}
   \includegraphics[width=12cm]{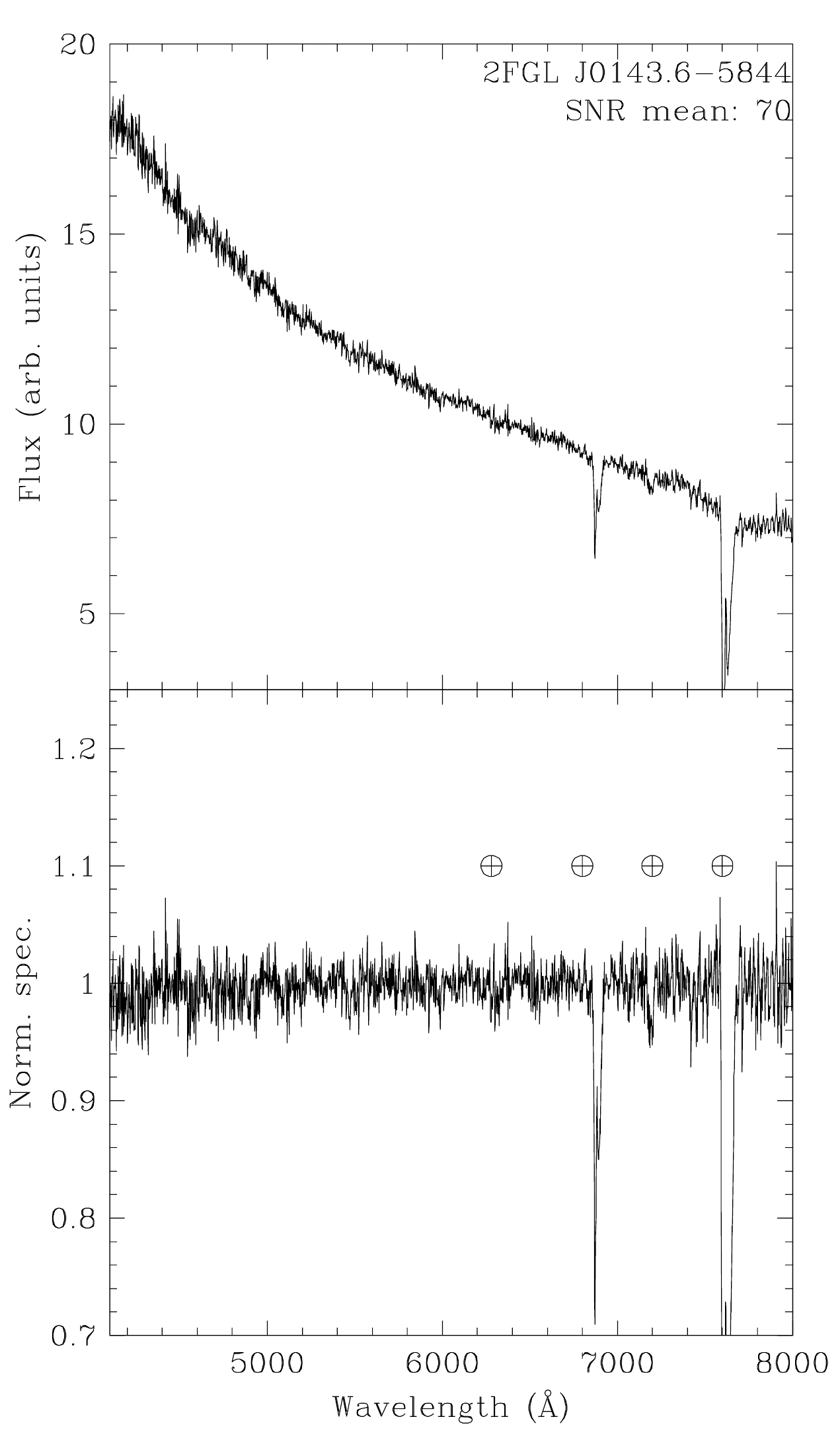}
     \caption{As in Figure \ref{fig:2fglj0115} but for 2FGL J0143.6-5844 .}
     \label{fig:2fglj0143}
\end{figure*}
\end{center}

\begin{center}
\begin{figure*}
   \includegraphics[width=12cm]{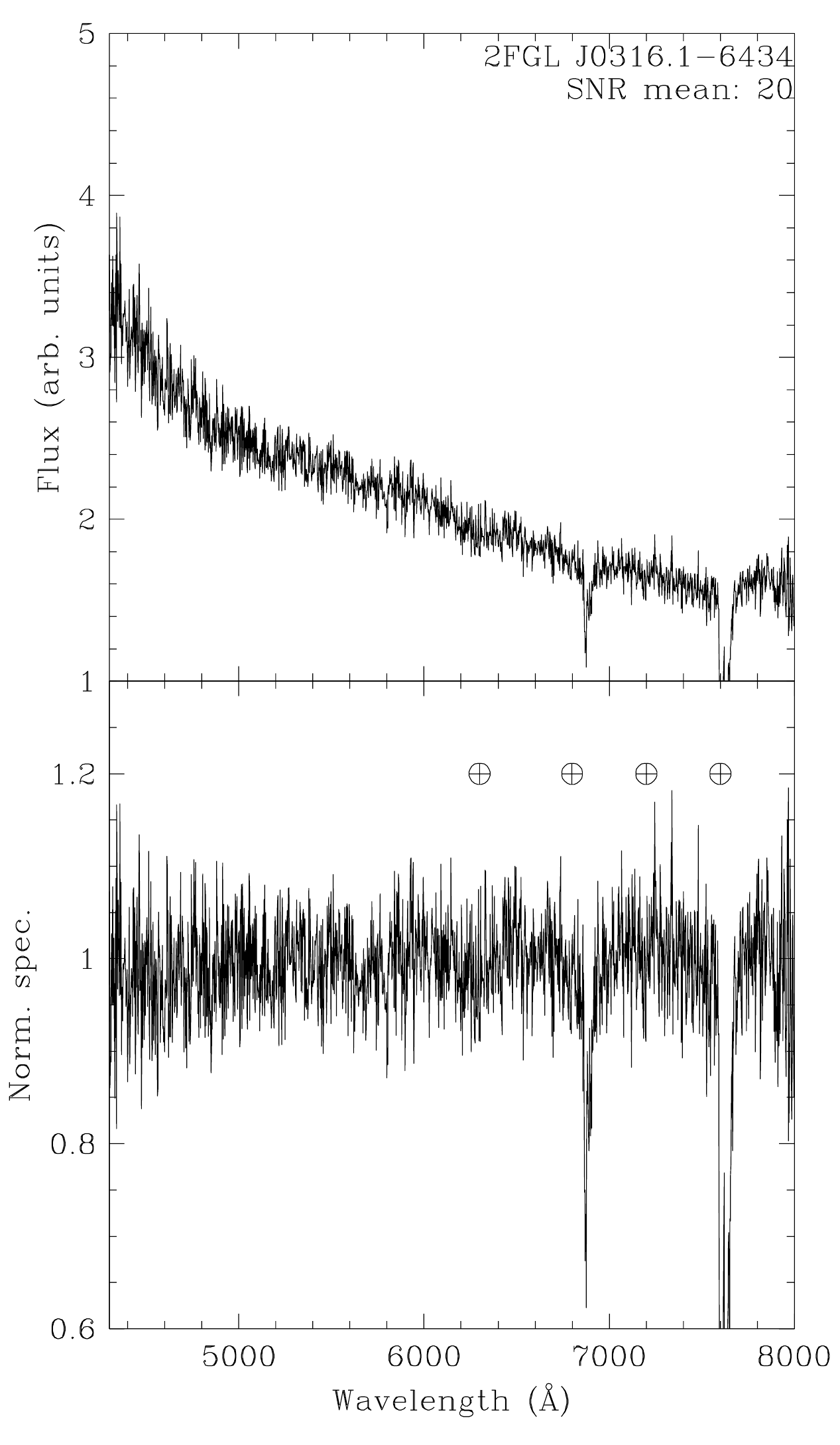}
     \caption{As in Figure \ref{fig:2fglj0115} but for 2FGL J0316.1-6434.}
     \label{fig:2fglj0316}
\end{figure*}
\end{center}

\begin{center}
\begin{figure*}
   \includegraphics[width=12cm]{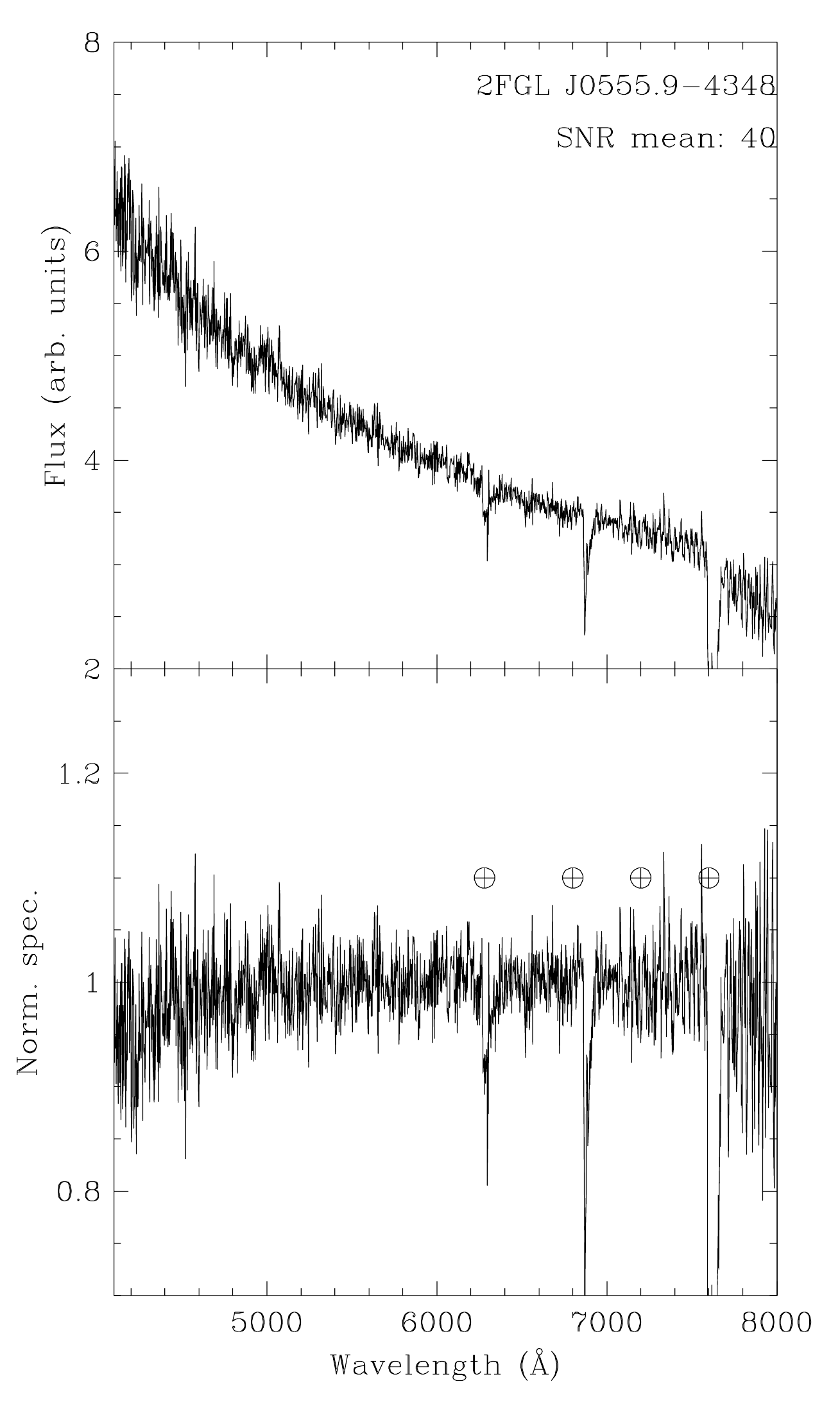}
     \caption{As in Figure \ref{fig:2fglj0115} but for 2FGL J0555.9-4348.}
     \label{fig:2fglj0555}
\end{figure*}
\end{center}

\begin{center}
\begin{figure*}
   \includegraphics[width=12cm]{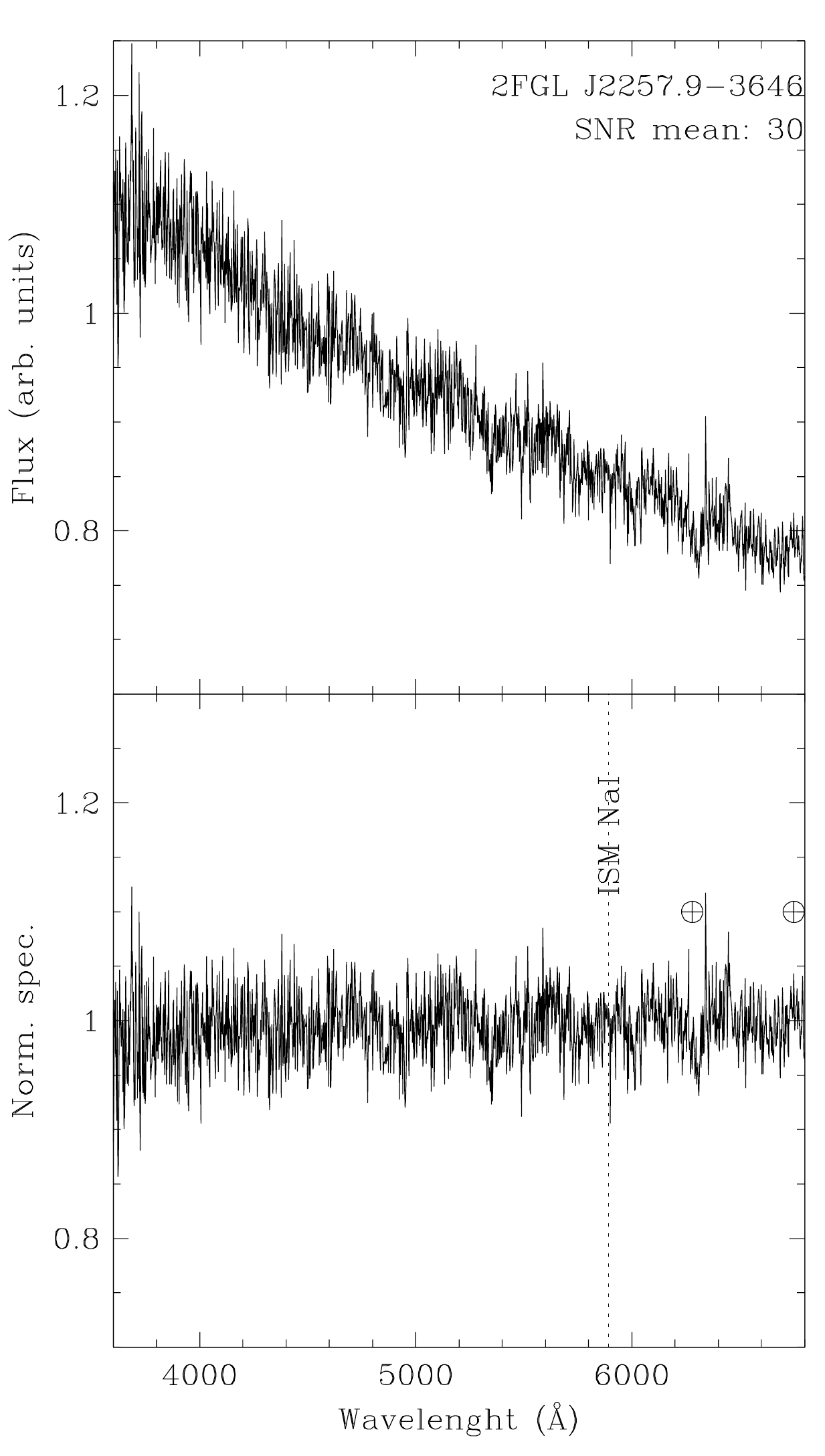}
     \caption{As in Figure \ref{fig:2fglj0115} but for 2FGL J2257.9-3646.}
     \label{fig:2fglj2257}
\end{figure*}
\end{center}

\begin{center}
\begin{figure*}
   \includegraphics[width=12cm]{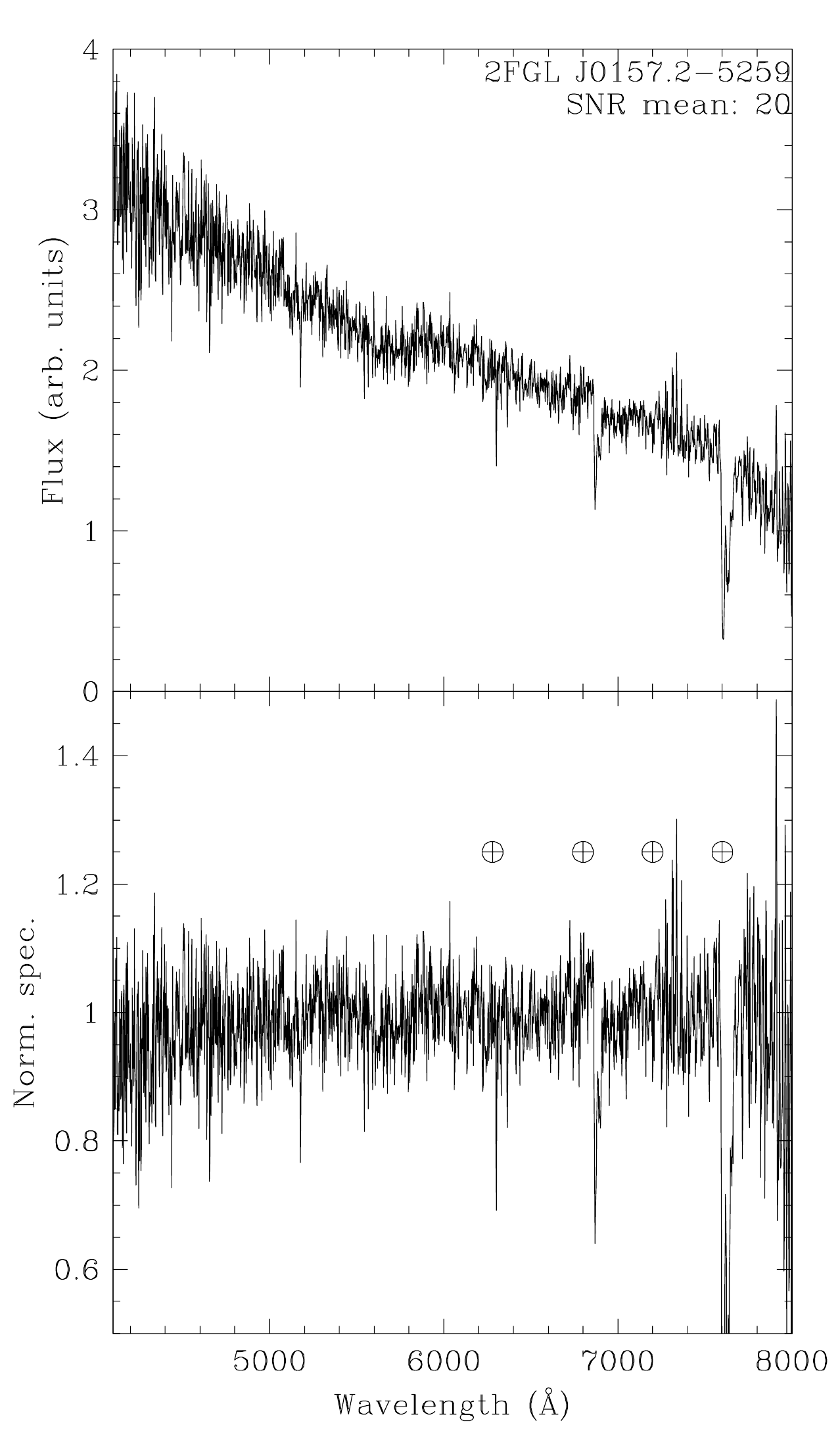}
     \caption{As in Figure \ref{fig:2fglj0115} but for 2FGL J0157.2-5259.}
     \label{fig:2fglj0157}
\end{figure*}
\end{center}

\begin{center}
\begin{figure*}
   \includegraphics[width=12cm]{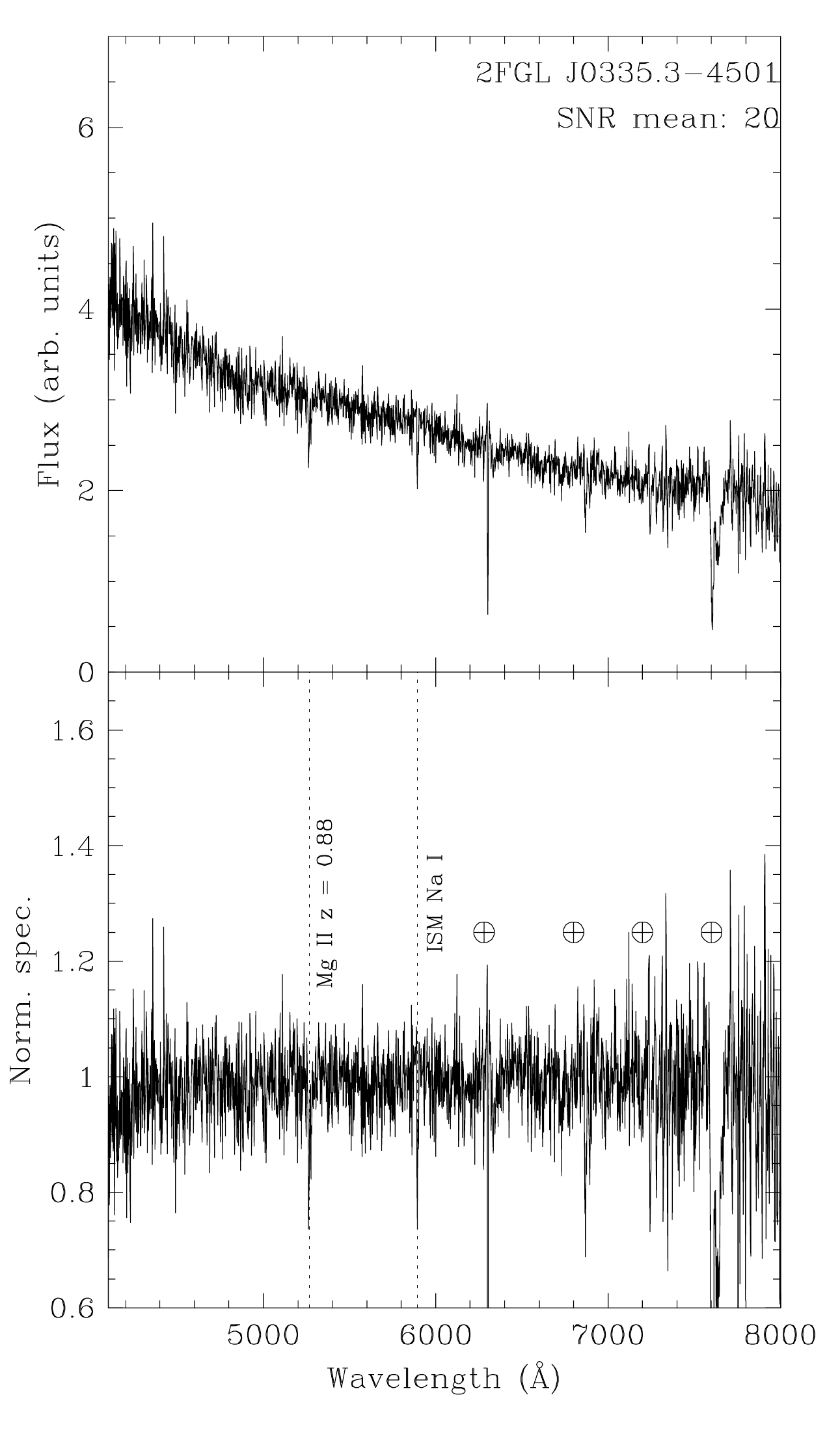}
     \caption{As in Figure \ref{fig:2fglj0115} but for 2FGL J0335.3-4501.}
     \label{fig:2fglj0335}
\end{figure*}
\end{center}

\begin{center}
\begin{figure*}
   \includegraphics[width=12cm]{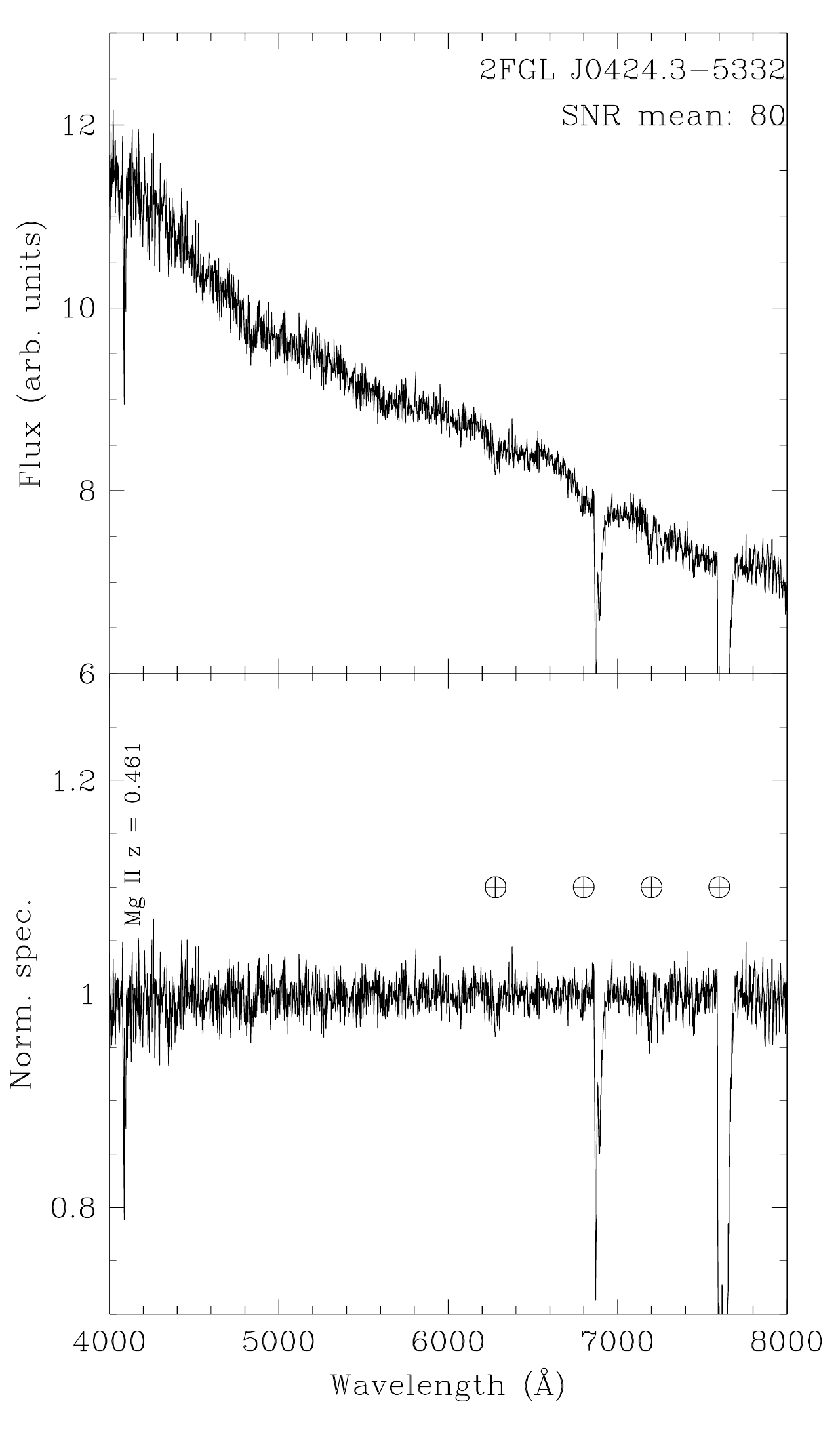}
     \caption{As in Figure \ref{fig:2fglj0115} but for 2FGL J0424.3-5332.}
     \label{fig:2fglj0424}
\end{figure*}
\end{center}

\begin{center}
\begin{figure*}
   \includegraphics[width=12cm]{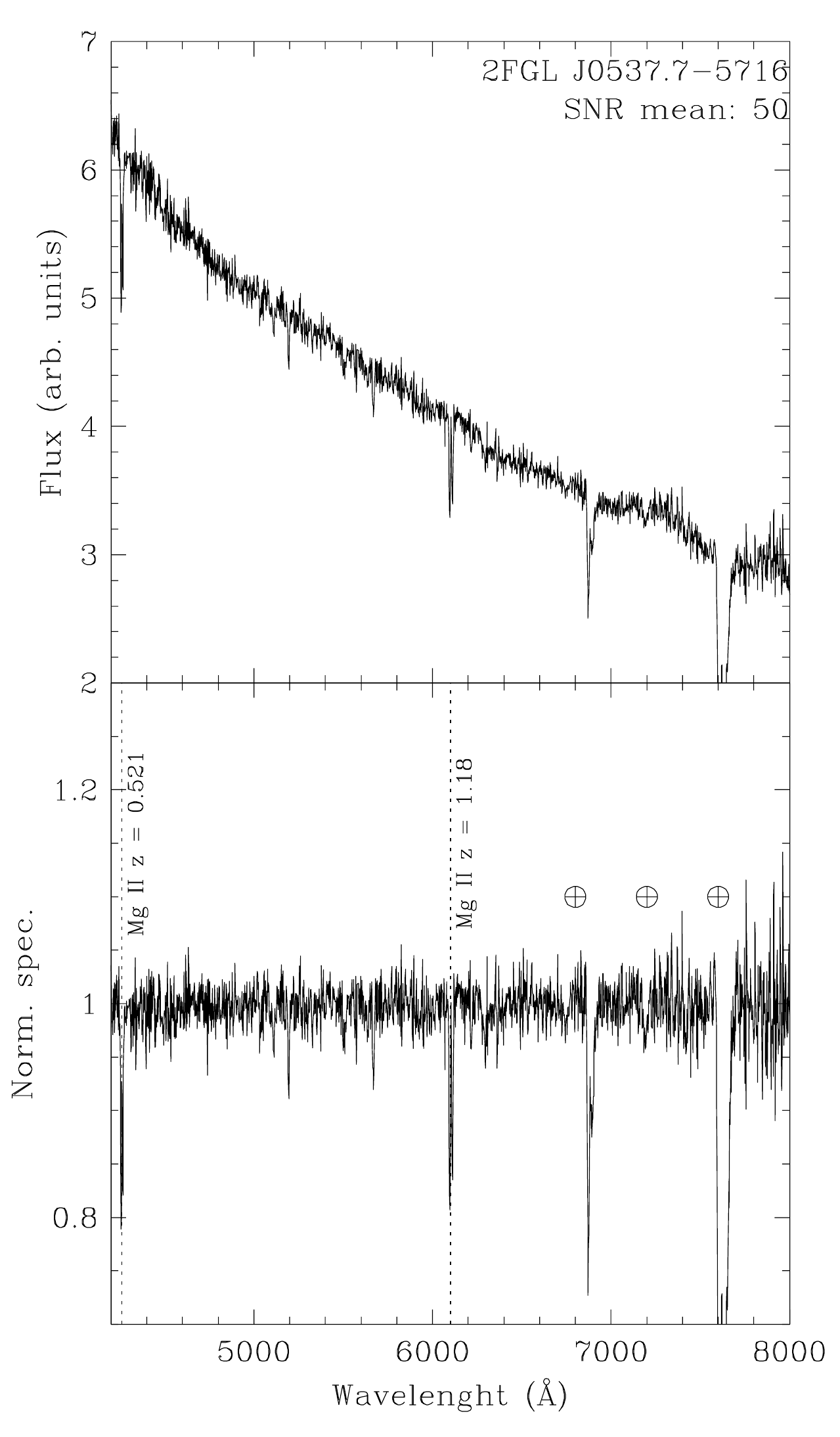}
     \caption{As in Figure \ref{fig:2fglj0115} but for 2FGL J0537.7-5716.}
     \label{fig:2fglj0537}
\end{figure*}
\end{center}

\begin{center}
\begin{figure*}
   \includegraphics[width=12cm]{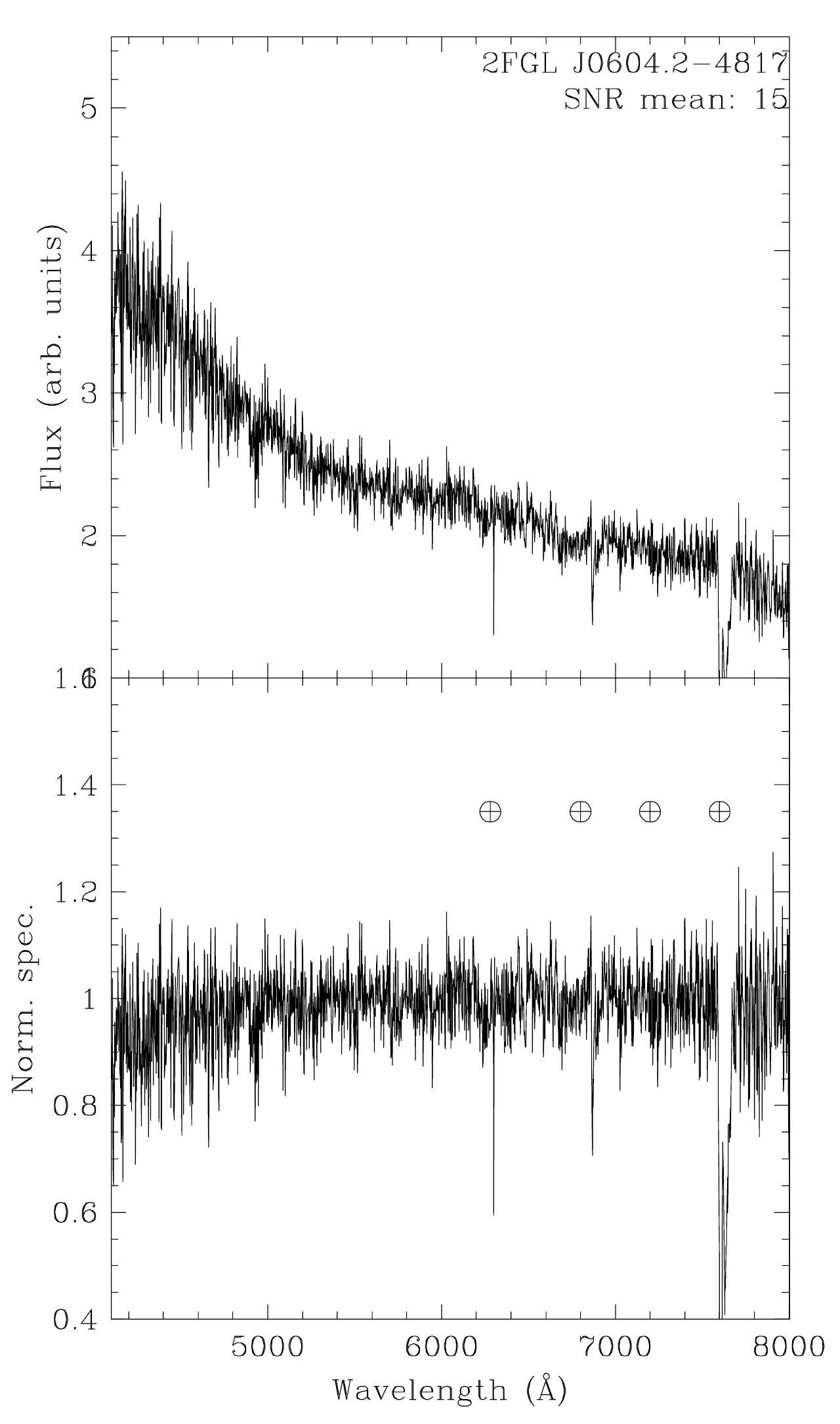}
     \caption{As in Figure \ref{fig:2fglj0115} but for 2FGL J0604.2-4817.}
     \label{fig:2fglj0604}
\end{figure*}
\end{center}

\begin{center}
\begin{figure*}
   \includegraphics[width=12cm]{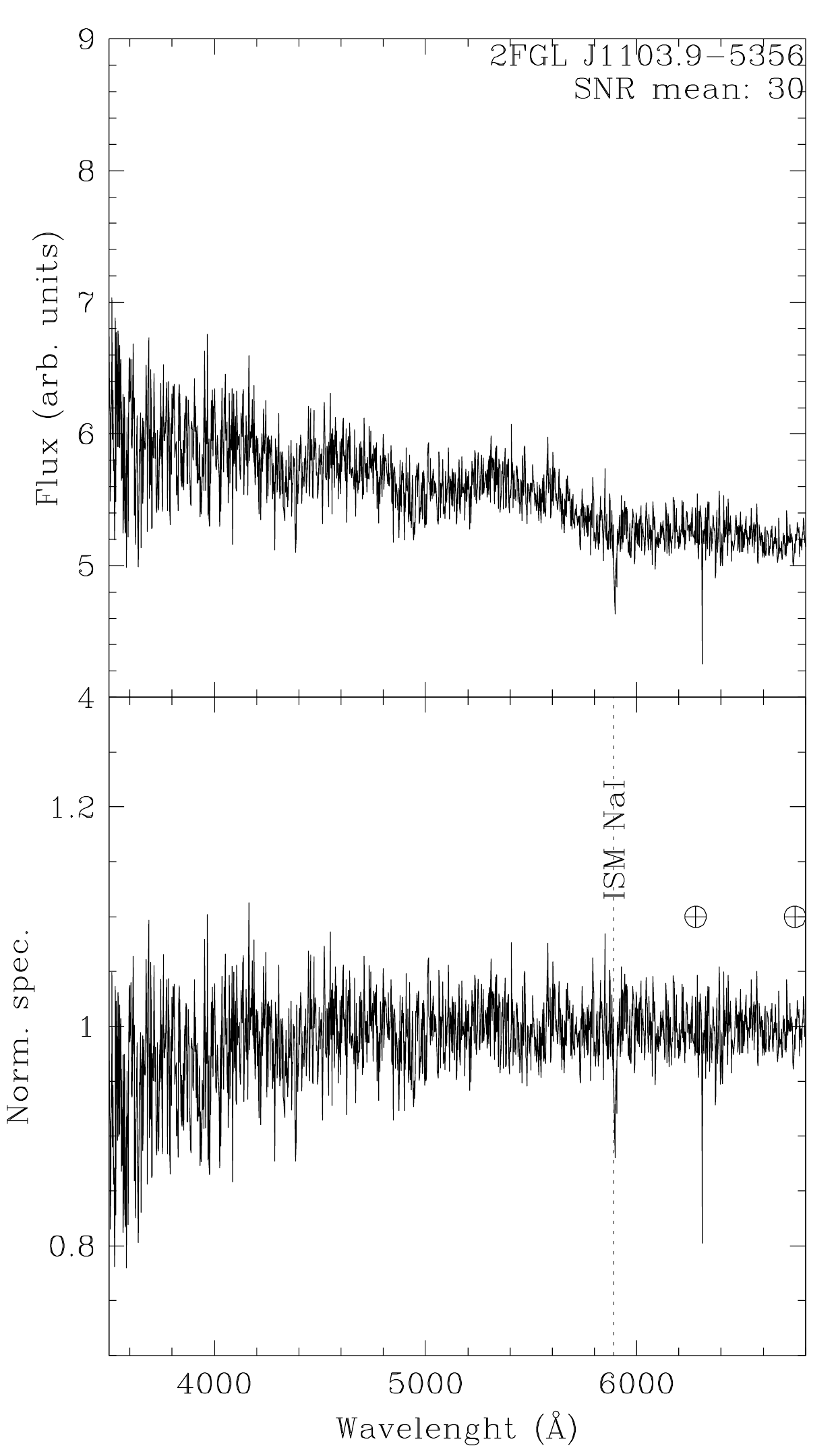}
     \caption{As in Figure \ref{fig:2fglj0115} but for 2FGL J1103.9-5356.}
     \label{fig:2fglj1103}
\end{figure*}
\end{center}

\begin{center}
\begin{figure*}
   \includegraphics[width=12cm]{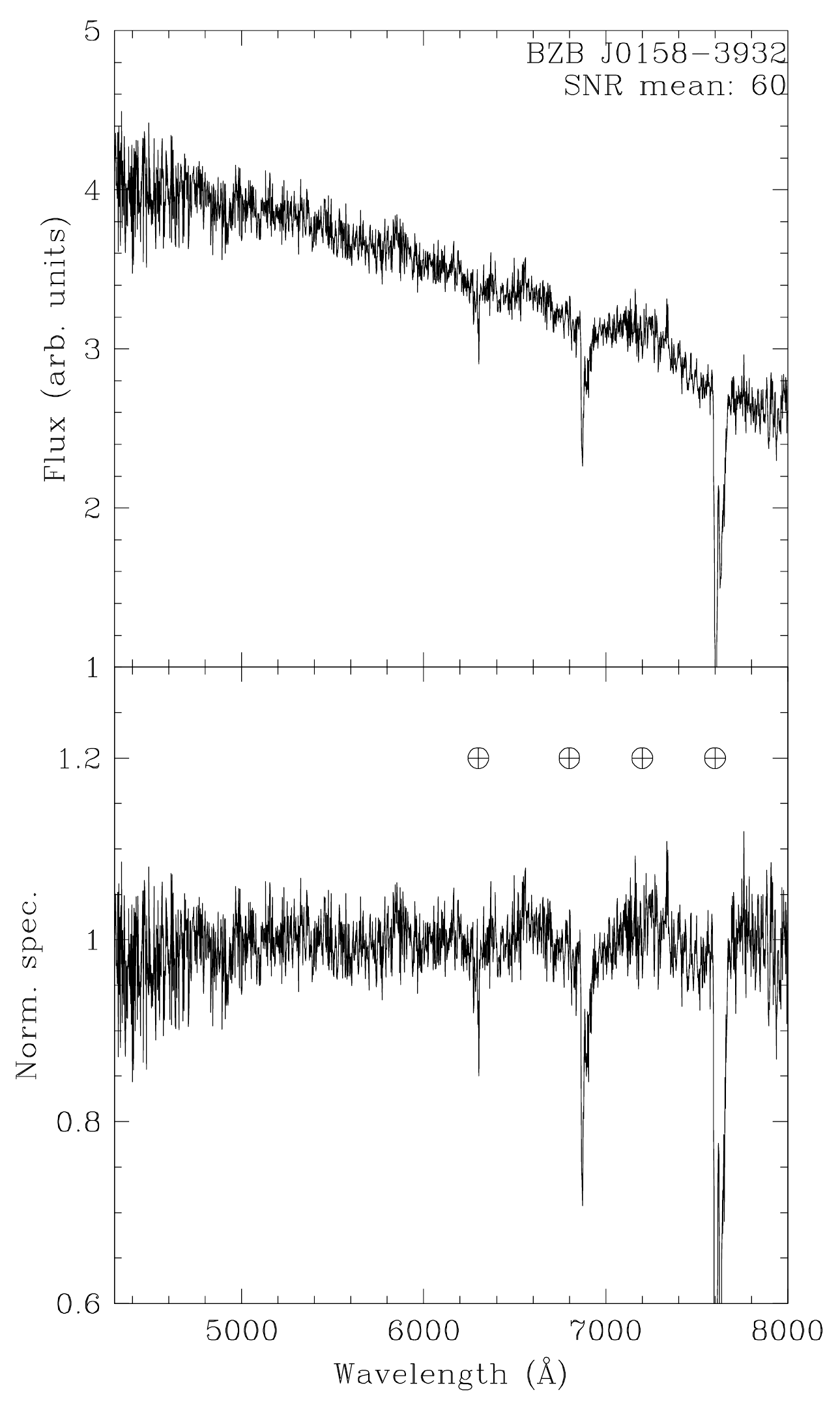}
     \caption{As in Figure \ref{fig:2fglj0115} but for BZB J0158-3932.}
     \label{fig:bzb0158}
\end{figure*}
\end{center}

\begin{center}
\begin{figure*}
   \includegraphics[width=12cm]{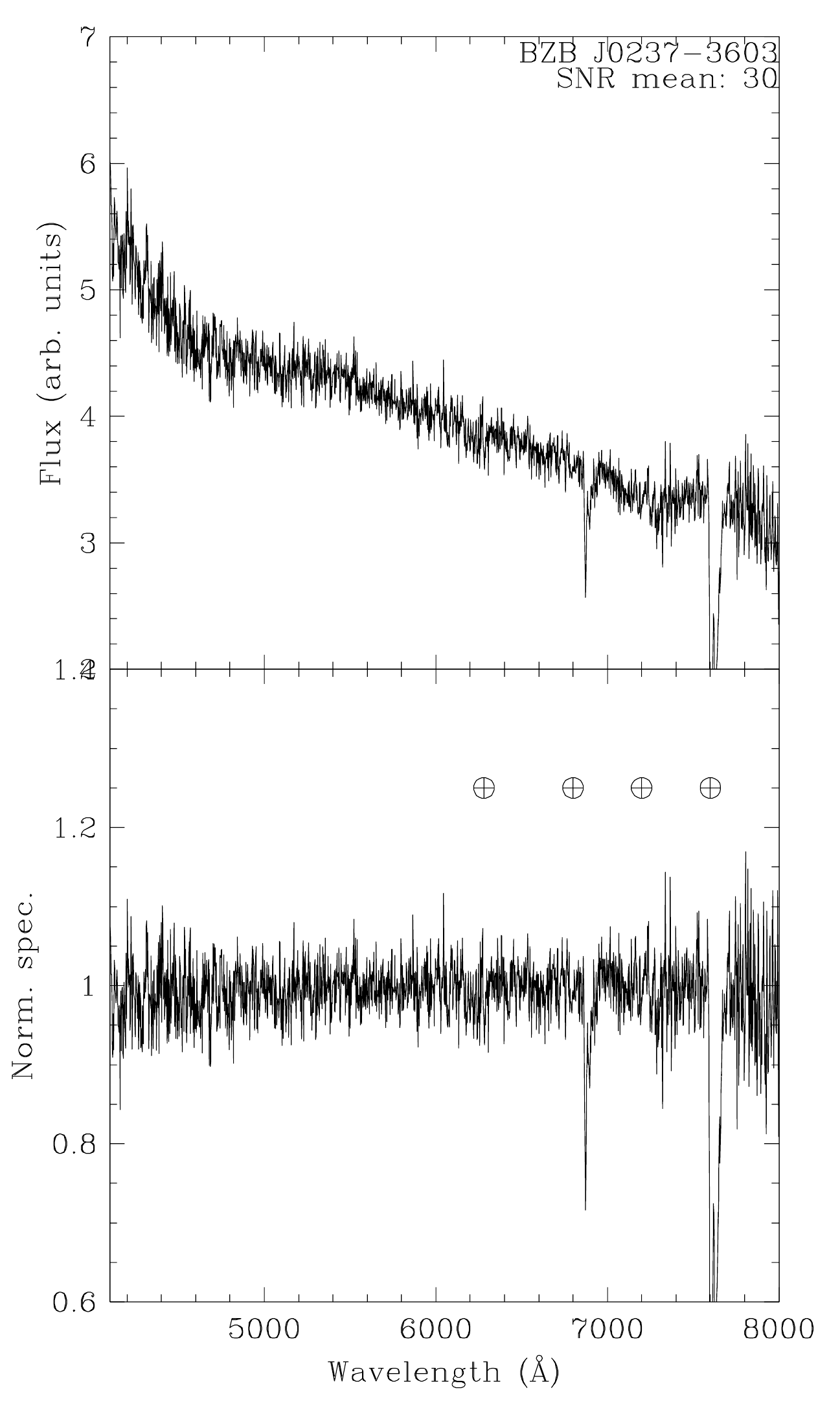}
     \caption{As in Figure \ref{fig:2fglj0115} but for BZB J0237-3603.}
     \label{fig:bzb0237}
\end{figure*}
\end{center}

\begin{center}
\begin{figure*}
   \includegraphics[width=12cm]{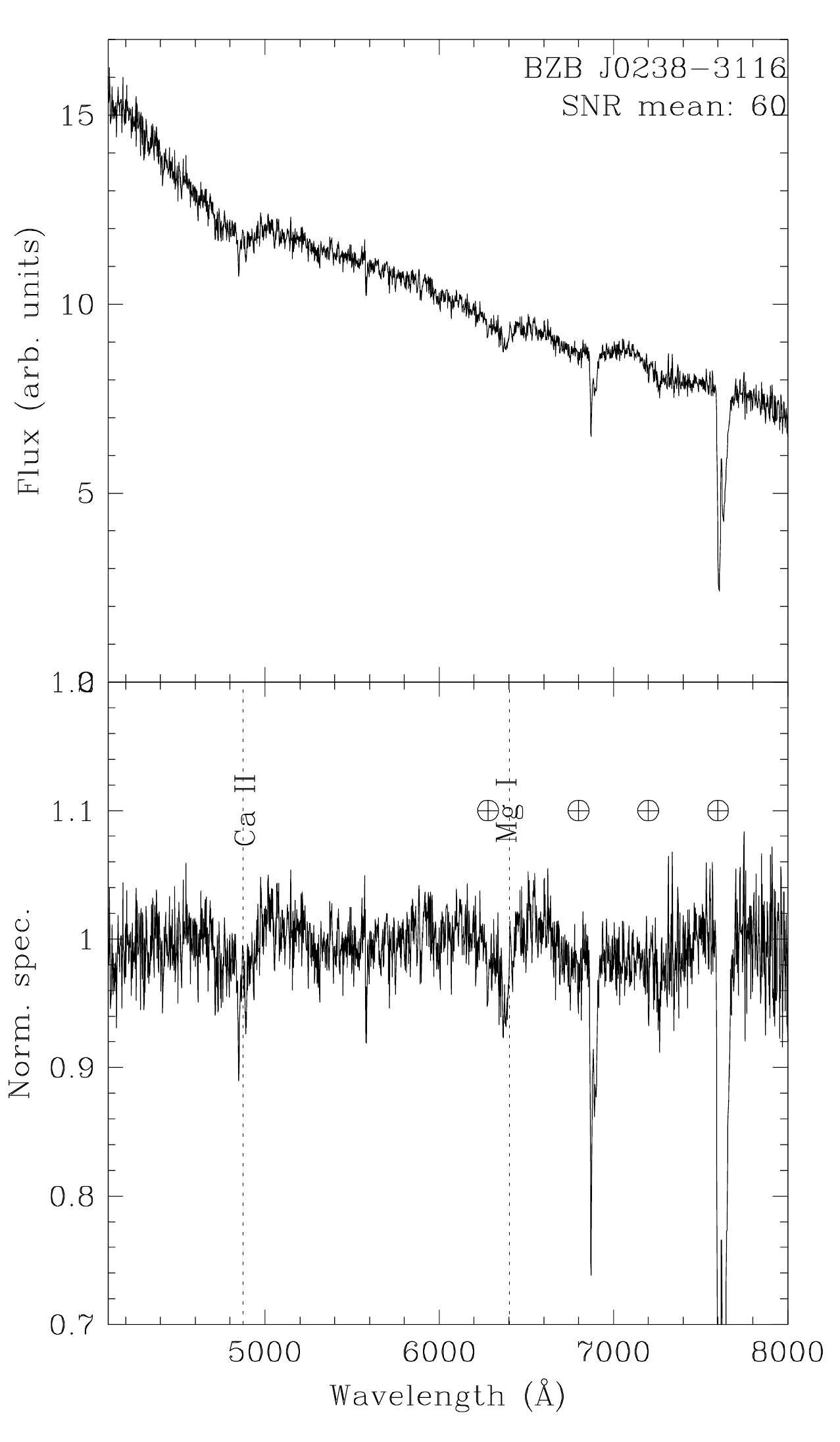}
     \caption{As in Figure \ref{fig:2fglj0115} but for BZB J0238-3116.}
     \label{fig:bzb0238}
\end{figure*}
\end{center}

\begin{center}
\begin{figure*}
   \includegraphics[width=12cm]{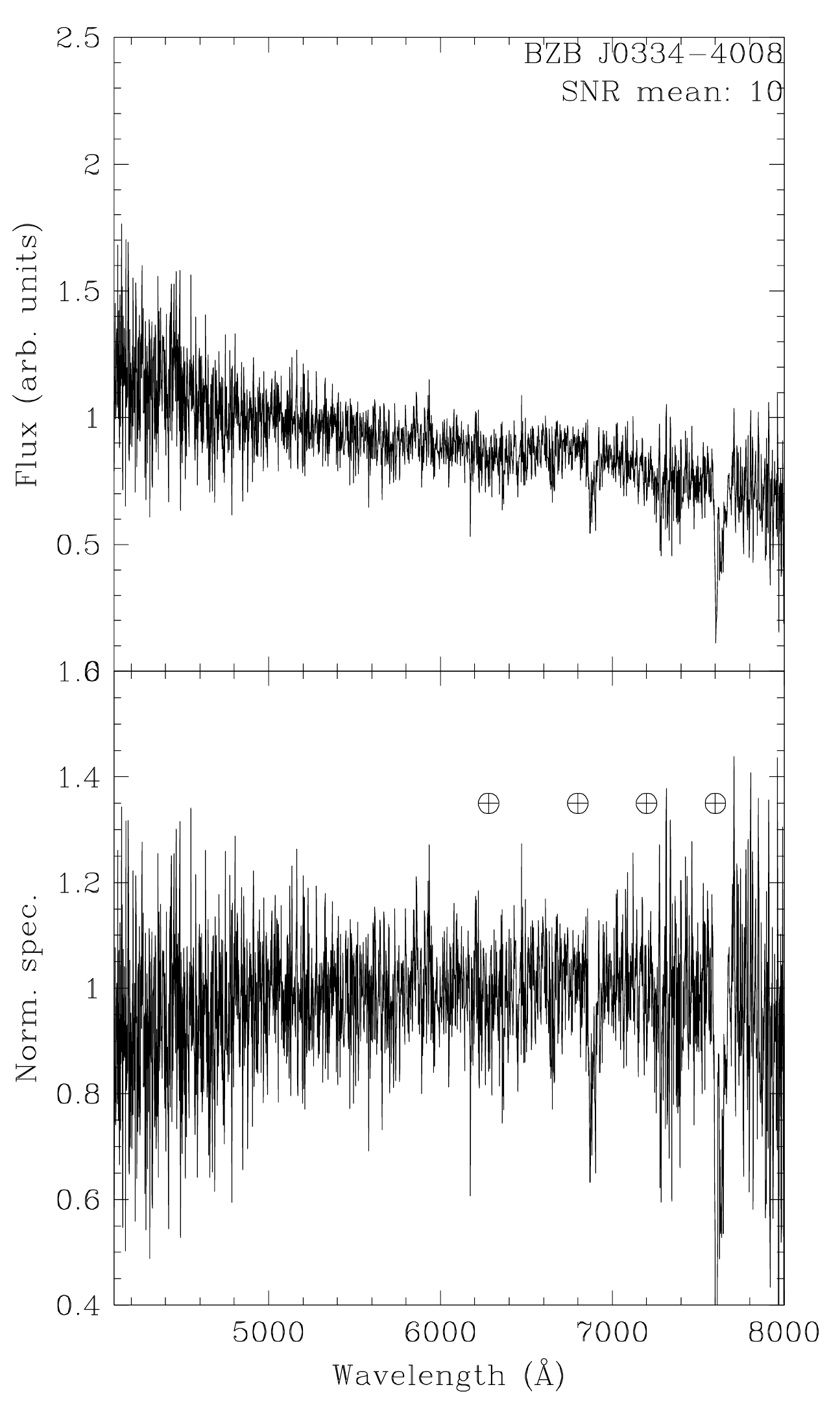}
     \caption{As in Figure \ref{fig:2fglj0115} but for BZB J0334-4008.}
     \label{fig:bzb0334}
\end{figure*}
\end{center}

\begin{center}
\begin{figure*}
   \includegraphics[width=12cm]{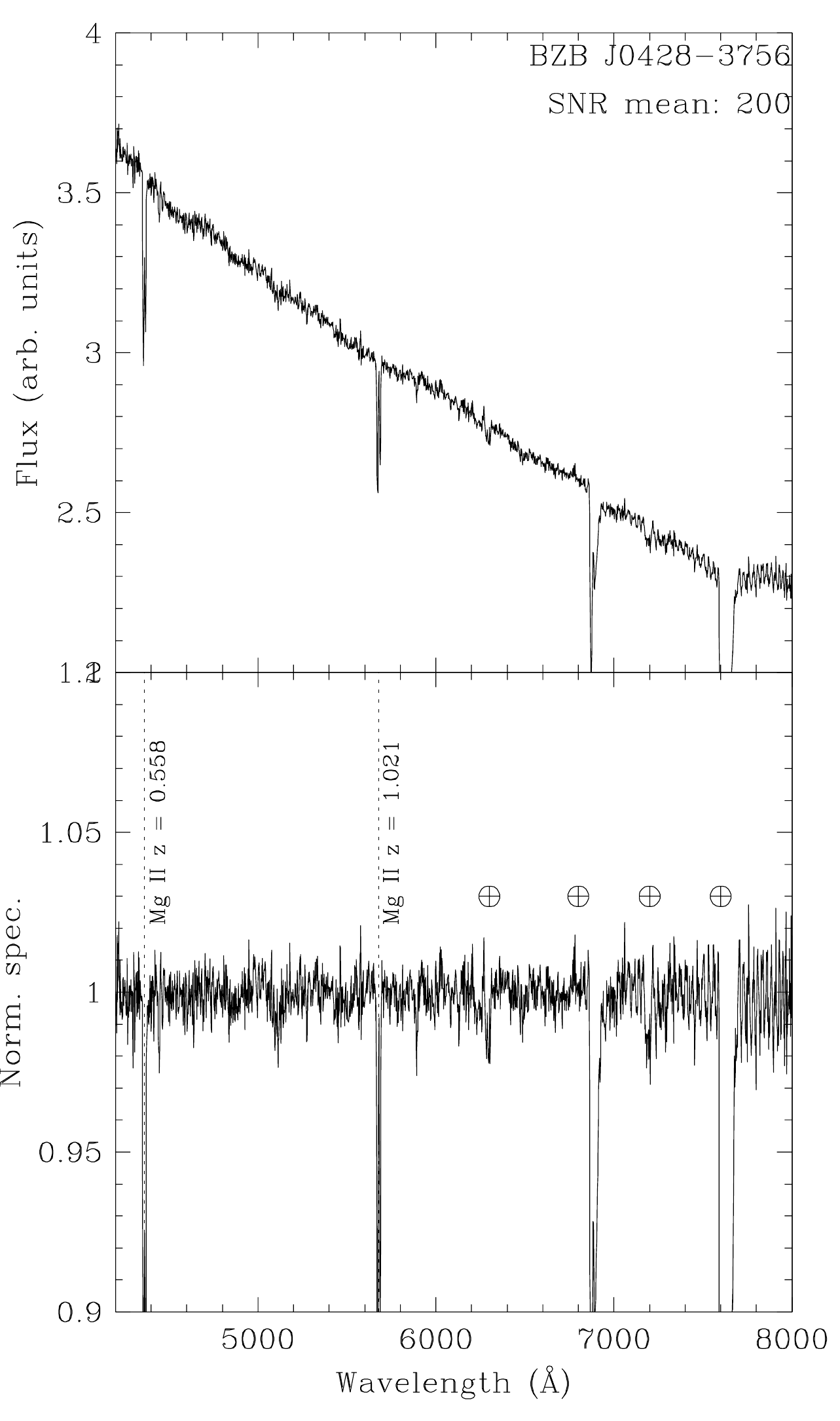}
     \caption{As in Figure \ref{fig:2fglj0115} but for BZB J0428-3756.}
     \label{fig:bzb0428}
\end{figure*}
\end{center}

\begin{center}
\begin{figure*}
   \includegraphics[width=12cm]{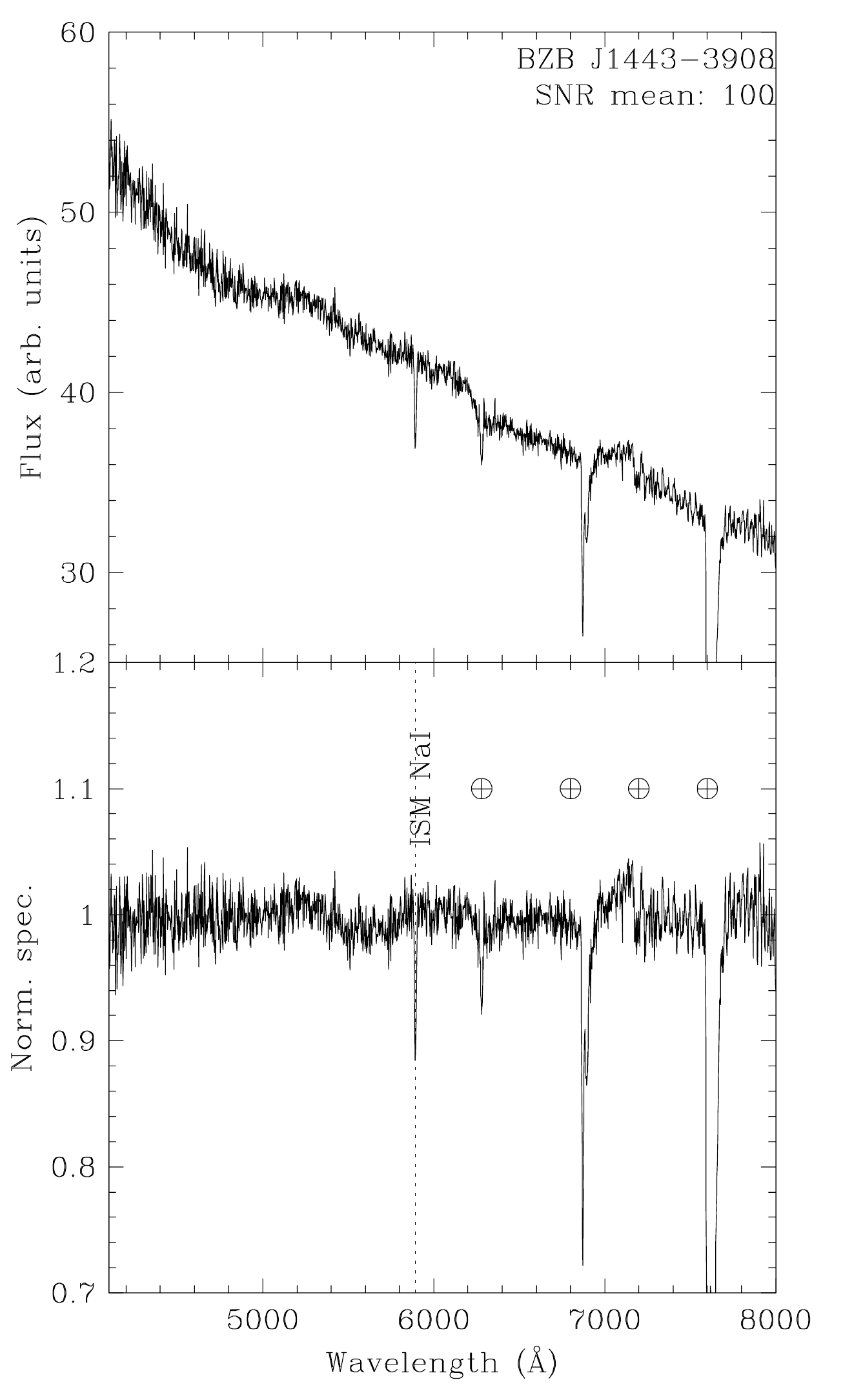}
     \caption{As in Figure \ref{fig:2fglj0115} but for BZB J1443-3908.}
     \label{fig:bzb1443}
\end{figure*}
\end{center}

We obtained spectra in Visiting mode at  the SOAR 4-m class telescope using the High Throughput Goodman spectrograph \citep{clemens04}. We adopted a slit of 1.0$^{\prime\prime}$ or 1.3$^{\prime\prime}$ (depending on the availability at the telescope) width combined with a low resolution grating (400 grooves mm$^{-1}$) yielding a dispersion of about 2 $\textrm{\AA}$ pixel$^{-1}$ ($\Delta{\lambda} \sim 5 \textrm{\AA}$) in both cases.
The majority of the spectra (14/19) were obtained in a dedicated run from September 15-16, 2013 while individual spectra were obtained for the remaining targets in August and November 2013 and January 2014 (see Table 1). The average seeing during both runs was about $1.0^{\prime\prime}$ and skies were almost clear.
Data reduction was done using IRAF\footnote{
IRAF (Image Reduction and Analysis Facility) is distributed by the National Optical Astronomy Observatories,
which are operated by the Association of Universities for
Research in Astronomy, Inc., under cooperative agreement
with the National Science Foundation.} by the adoption of standard procedures. In particular, for each object we performed bias subtraction, flat field and cosmic rays rejection. Since for each target we secured 2 individual frames, we averaged them according to their signal to noise ratios. {We rejected any spurious features (e.g cosmic rays or CCD defect) by comparing the two individual exposures.}
The wavelength calibration was achieved using the spectra of Helium-Neon-Argon or Iron-Argon lamps which assures a full coverage of the entire range. In order to take into account flexures of the instruments and drift due to poor long term stability during the night, we took an arc frame before each target in order to guarantee a good wavelength calibration for the scientific spectra. The accuracy achieved is about 0.2 $\textrm{\AA}$ root mean square (rms).
We observed a photometric standard star each night, even though this program does not require an accurate flux calibration, in order to ensure a flux calibration of spectra. We applied a correction law for Galactic reddening \citep{card89} assuming $E_{B - V}$ values computed by Schlegel et al. 1998.
Finally, in order to better investigate the detectability of faint spectral features, we normalised each spectrum by dividing by the continuum best fit computed on the observed data. We report in Figure \ref{fig:fc1}, \ref{fig:fc2} and \ref{fig:fc3} the finding charts of each object while the full spectra are reported from Figure \ref{fig:2fglj0115} to Figure \ref{fig:bzb1443}. Full spectra will be also public available at the Web site \texttt{http://www.oapd.inaf.it/zbllac/} \citep{sbazbllac}.

\section{Results}
\label{sec:results}
{According to the optical spectra obtained during our campaign, we found that all the sources selected as counterparts of the seven UGSs are indeed blazars (see Section \ref{sec:ugs}) . The six AGUs, classified according to the Second \fer-LAT AGN catalog \citep[2LAC;][]{ackermann11a} are also BL Lacs that show typical power-law spectrum with weak intrinsic (or absent) emission features with rest frame equivalent width EW $\geq$ 5$\textrm{\AA}$ (see Section \ref{sec:gamma}).  
The remaining six sources were labeled as BL Lac candidates in the Roma BZCAT because no optical spectra were found can now be classified as BL Lac from careful inspection of each spectrum secured in this campaign (see Section \ref{sec:bzcat}).} %Source details are given in the next sections. 
{Although some of the sources considered in our campaign have been also observed at different observatories and groups, as reported in the following sections, we decided to reobserve these targets for two main reasons. First, at the time when our observations were scheduled and performed these spectra were not yet published and, second, to the well-know BL Lac variability in the optical band there is always the chance to observe the source in a low state and detect some  features allowing the determination of their redshifts. Moreover, for a number of objects the available spectra in the literature exhibit rather low signal to noise ratios making it almost impossible to secure an affordable classification of sources and, if possible, the determination of their distance. Source details are given in the next section. }

\subsection{Unidentified Gamma-ray Sources (UGS)}
\label{sec:ugs}
{The \wse\ selected counterparts to these UGSs show IR \wse\ colors similar to those of $\gamma$-ray blazars considered in the analysis performed by Massaro et al. (2013a). In particular, they were predicted to be BL Lacs rather than FSRQs \citep[see][]{ugs1} since their IR colors are consistent with those of the \fer\ BL Lacs. 
Such probabilites are computed adopting the same procedures described in the previous Fermi catalog releases (add refs. x 1FGL and the 2FGL). 
They are all detected in the radio band, having counterparts in the Sydney University Molonglo Sky Survey \citep[SUMSS;][- S]{mauch03} in agreement with the expectations of the radio-$\gamma$-ray connection \citep[e.g.,][]{ghirlanda10,mahony10,ugs3}.
Five out of seven sources are detected in the Two Micron All Sky Survey \citep[2MASS;][- M]{skrutskie06}
while the whole sample of UGSs have an optical counterpart in the USNO-B1 Catalog \citep[][- U]{monet03}.%as shown in the finding charts
%(see Figure~\ref{fig:fc1} and Figure \ref{fig:fc2}).
 We report the main multifrequency properties of each source in Table~\ref{tab:log}, together with their 2FGL name,
the \wse\ name, and the counterpart coordinates. We also searched for optical counterparts in the Six-degree-Field Galaxy Survey \citep[6dFGS;][- 6]{jones04,jones09} while, at high energies, we analysed the ROSAT all-sky survey  in both the ROSAT Bright Source Catalog \citep[RBSC;][- X]{voges99}  and the ROSAT Faint Source Catalog \citep[RFSC;][- X]{voges00} as well the Deep Swift X-Ray Telescope Point Source Catalog \citep[1SXPS;][- x]{evans14} .We also took into account the \swf\ X-ray survey for all the \fer\ UGSs$^9$ \citep{stroh13,paggi13,takeuchi13} \footnote{\underline{http://www.swift.psu.edu/unassociated/}}. We adopt the same symbol for the X-ray catalog of \xmm\ , \chn\ and \swf\ because they provide observations for \textit{only pointed} source while most of the X-ray counterparts of our potential counterparts can be serendipitous.}

\begin{center}
\begin{figure*}
   \includegraphics[width=12cm]{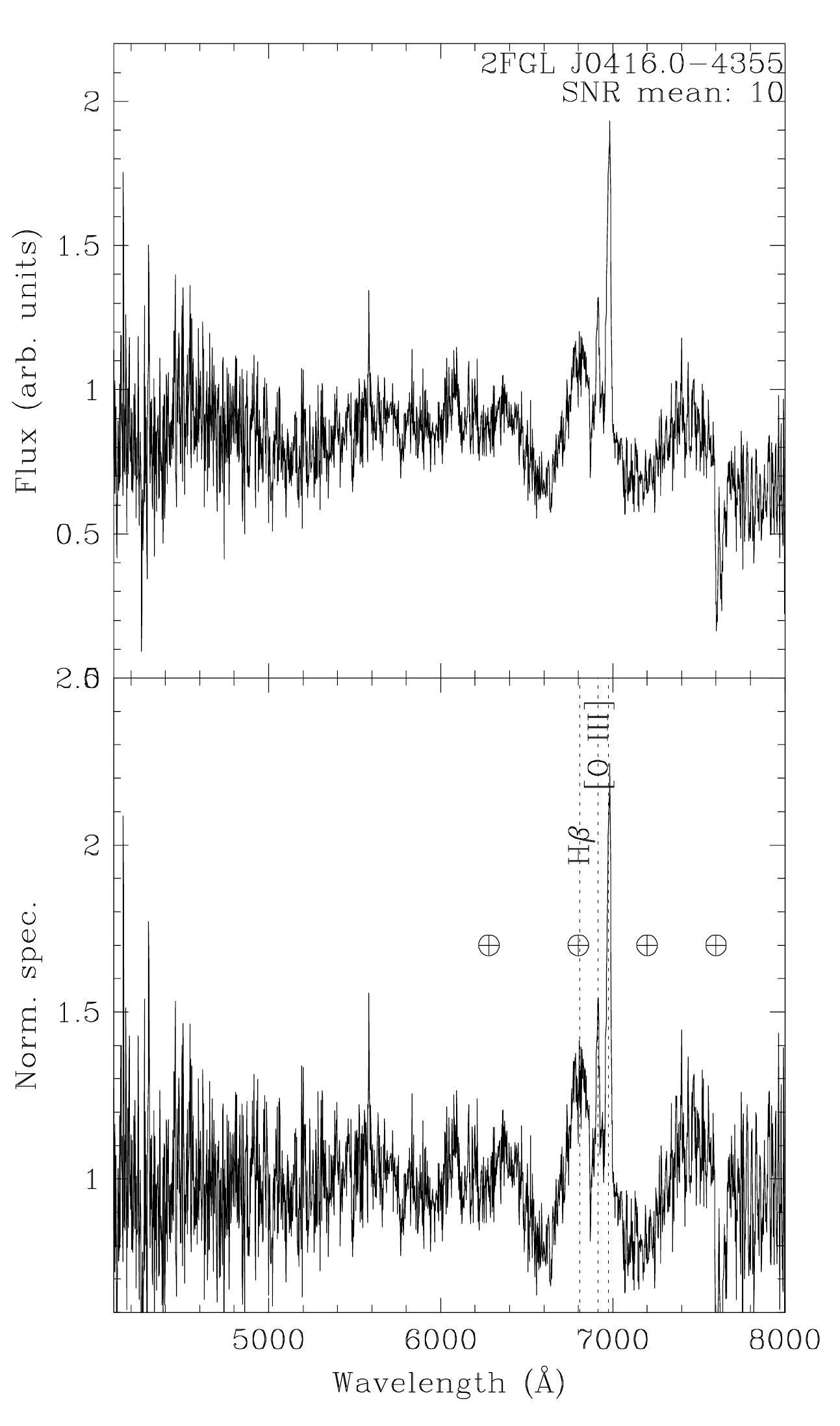}
     \caption{As in Figure \ref{fig:2fglj0115} but for 2FGL J0416.0-4355.}
     \label{fig:2fglj0416}
\end{figure*}
\end{center}

In agreement with the expectations of the method developed by D'Abrusco et. al. 2013,
for six out of seven sources our spectroscopic observations confirmed their BL Lac nature (see Figures \ref{fig:2fglj0115} to \ref{fig:2fglj2257}). We note that their completely featureless spectra do not allowed us to determine their redshifts.
The remaining one, namely 2FGL J0416.0-4355, appears to have a quasar-like spectrum in the optical band (see Figure\ref{fig:2fglj0416}),
similar to FSRQ sources with a redshift estimate of 0.398 from H$\beta$ ($\lambda$ 6800, EW 110$\textrm{\AA}$) and [O III] doublet ($\lambda\lambda$ 6913-6976, EW 25-50 $\textrm{\AA}$).  However, in this case the lack of additional information necessary to compute the radio spectral index did not permit us to
classify the source as FSRQ. {Finally, we highlight that the chance probabilities of having a spurious association for the WISE sources selected according to their blazar-like IR colors and confirmed as blazars thanks to these new observations will be assessed in the upcoming Fermi-LAT catalog release once these new blazars are added to comparison catalogs of potential counterparts (Abdo et al. 2015)}

\subsection{Gamma-ray Active Galaxies of Uncertain type (AGU)}
\label{sec:gamma}
%We report in Table~\ref{tab:log} the multifrequency notes relative to each AGU, as well as for the UGSs previously described. 
{In the sample of AGUs, chosen on the same basis of the IR colors analysis performed for UGSs, three out of six sources
belong to the First \fer-(LAT  Source Catalog \citep[1FGL;][]{abdo10}, as marked in Table~\ref{tab:log}, meanwhile
all the AGUs are detected in the X-ray and show counterparts 
in both the RBSC and RFSC, as well as radio counterparts in the SUMSS.}
In particular, 2FGL J0424.3-5332 is also detected in the Parkes-MIT-NRAO Surveys \citep[PMN;][- Pm]{wright94}
as well as in the Australia Telescope 20 GHz Survey \citep[AT20G;][- A]{murphy10} and, since it has a flat radio spectrum,
also belong to the Combined Radio All-Sky Targeted Eight-GHz Survey \citep[CRATES;][- c]{healey07}.
2FGL J0157.2-5259 is also detected by GALEX (NASA Extragalactic Database (NED)\footnote{\underline{http://ned.ipac.caltech.edu/}}).

As shown by \cite{paper3}, four of the AGUs have \wse\ colors typical of \fer\ blazars at 95\% level of confidence,
while the other two, namely 2FGL J0335.3-4501 and 2FGL J0604.2-4817\footnote{Also observed during our program even if its optical spectrum was already published in Masetti et al. (2013) since our data were taken while that paper was submitted.},  are only marginally consistent with the \wse\ Gamma-ray Strip. %\citep{}.
However they are all detected by \wse\ and five out of six also have counterparts in 2MASS. Moreover, the X-ray, $\gamma$-ray, and radio properties of the WISE counterparts of 2FGL J0157.2-5259 (Takeuchi et al. 2013) and 2FGL J1103.9-5356 \cite{kei11} are also consistent with the blazar nature for these sources.

Our spectroscopic observations confirm that all six sources are BL Lac objects (see Figure \ref{fig:2fglj0157} to \ref{fig:2fglj1103}).
For three of them, 2FGL J0335.3-4501, 2FGL J0424.3-5332 and  2FGL J0537.7-5716, we have been able to set a lower limit on their 
redshifts based on the detection of Mg II absorption lines that could be due to the blazar host galaxy 
or, more probably, to intervening systems. The lower limit on the redshift is $z \geq 0.88$ for 2FGL J0335.3-4501 and $z \geq 0.461$ for 2FGL J0424.3-5332, respectively. More interesting is the case of 2FGL J0537.7-5716 where multiple Mg II absorption system are detected. The first one is found to be at redshift $z = 0.521$ while the second one is at $z = 1.18$ making this source one of the most distant and infrared luminous ($L \geq 3\times10^{45} \textrm{erg } \textrm{s}^{-1}$) BL Lac ever observed. Finally, we note that the sources 2FGL J0424.3-5332 and 2FGL J1103.9-5356 also appear in Shaw et. al 2013a. but for the first one we are able to set a spectroscopic lower limit to the redshift from an intervening system of Mg II ($z = 0.461$, in agreement with the proposed lower limit) that it was not detected by Shaw et. al 2013a.  while, for the second source, our spectrum exhibits an higher SNR with respect to the one reported in the same paper.

\subsection{$\gamma$-ray blazars}
\label{sec:bzcat}
We report in Table~\ref{tab:log} for our 6 $\gamma$-ray blazars their \bzcat\ and \wse\ name.
No multifrequency notes are present in this table since they are already discussed in the \bzcat.
All these 6 blazars belong to the sample named {\it locus} used in D'Abrusco et. al. 2013 to identify blazar-like sources within the positional uncertainty region of the \fer\ UGSs.
Thus all the \wse\ counterparts have the IR consistent with the remaining \fer\ blazars.

The first four blazars listed in Table~\ref{tab:log} are BL Lac candidates, thus no optical spectrum was available in literature
preventing us to provied a firm classification. The remaining two, BZB J0428-3756 and BZB J1443-3908, 
were indeed confirmed blazars with an uncertain redshift estimate.
Our spectroscopic observations verified that all the BL Lac candidates have featureless optical spectra
as shown in Figures ~\ref{fig:bzb0158}-~\ref{fig:bzb1443} with the only exception of BZB J0238-3116 for which the
CaII ($\lambda\lambda$ 4850-4891, EW $\sim$ 1.00-0.70$\textrm{\AA}$) and Mg I ($\lambda$ 6420, EW $\sim$ 2.50$\textrm{\AA}$) absorption lines from its host galaxy starlight allow us to estimate the redshift $z = 0.232$. We finally note that in the spectrum of BZB J0428-3756 multiple Mg II intervening systems ($z = 0.558$ and $z = 1.02$) are clearly detected allowing us to put a stringent spectroscopic lower limit to this high redshift BL Lac object. These absorption features are also confirmed from spectroscopic study of \cite{heidt}. However, we could not detect the broad emission feature at 5906$\textrm{\AA}$ ascribed to Mg II as reported in \cite{heidt} suggesting that our observation may have been carried out while the source was in a high state.
\\

We note that three of four sources (BZB J0158-3932, BZB J0334.2-4008 and BZB J1443.9-3908) appear also in Shaw et. al. 2013a. Nevertheless, while for BZB J0158-3932 and BZB J1443.9-3908 the spectra are similar in terms of SNR and spectral coverage, for BZB J0334.2-4008 we could not confirm the detection of broad emission lines of Mg II and C III] reported in Shaw et. al. 2013a supporting the fact that we probably observed the source during a high state of the relativistic jet which outshined the underlying emission lines from the BLR.

\section{Summary and conclusions}
\label{sec:conclusions}
We report the results of our exploratory spectroscopic campaign 
carried out in the Southern hemisphere with the SOAR telescope from August 2013 to January 2014.
We observed a selected sample of 19 targets.
Seven sources were \wse\ counterparts of \fer\ UGSs listed in the 2FGL
and identified thanks to the IR-based procedure developed by D'Abrusco et. al. 2013 and Massaro et. al. 2013a
In addition we collect spectra for 6 AGUs associated in the 2FGL catalog and 6 $\gamma$-ray BL Lac candidates
for which no optical spectra where found in literature leaving the classification uncertain.

{
Our main goal was to confirm the nature of the counterparts for the UGSs selected on the basis of their IR colors and correctly classify the remaining targets. 
We found that 6 out of 7 \wse\ sources chosen to be potential counterparts of the UGSs are clearly BL Lac objects 
while 2FGL J0416.0-4355 appears to be identified with a QSO at $z = 0.398$. The lack of radio information did not allow us to classify this source as a FSRQ.
These results are fully in agreement with the predictions of our developed association methods \citep[see][for more details]{ugs2}.
We also discovered  that all the AGUs observed are indeed BL Lac objects and for two of them we were able to report a first redshift estimate
employing Mg II intervening absorption system along the line of sight (or due to their host galaxies).
Finally, all the BL Lac candidates that belong to the \wse\ sample of \fer\ blazars used in the sample adopted to calibrate the association methods (e.g. \citep{ugs1,paper2}) are firmly established as bona fide BL Lacs through our spectroscopic observations.
For BZBJ0238-3116 we measured a redshift $z = 0.232$ through Ca II break and Mg I absorption lines from host galaxy starlight, while for BZBJ0428-3756 we put a lower limit to the redshift at $z \geq 1.02$ from an intervening Mg II absorption system detected along the line of sight.
}
%%WISE FINDING CHART

%width = 12cm
%----------    UGS ----------

%---- AGU

%%--Roma BZCAT Sources

%\begin{appendix}

\acknowledgements
% grants
This investigation is supported by the NASA grants NNX12AO97G and NNX13AP20G.
H.A.S. acknowledges partial support from NASA/JPL grant RSAs 1369566 , 1369556 and 1369565. The work by G. Tosti is supported by the ASI/INAF contract I/005/12/0 while work by C.C.C. at NRL is supported in part by NASA DPR S-15633-Y. 
We are grateful to Dr. S. Points for his help to schedule, prepare and perform the SOAR observations
% ASDC
Part of this work is based on archival data, software or on-line services provided by the ASI Science Data Center.
% HEASARC
This research has made use of data obtained from the high-energy Astrophysics Science Archive
Research Center (HEASARC) provided by NASA's Goddard Space Flight Center; 
% SIMBAD and NED
the SIMBAD database operated at CDS,
Strasbourg, France; the NASA/IPAC Extragalactic Database
(NED) operated by the Jet Propulsion Laboratory, California
Institute of Technology, under contract with the National Aeronautics and Space Administration.
% SUMSS
The Molonglo Observatory site manager, Duncan Campbell-Wilson, and the staff, Jeff Webb, Michael White and John Barry, 
are responsible for the smooth operation of Molonglo Observatory Synthesis Telescope (MOST) and the day-to-day observing programme of SUMSS. 
The SUMSS survey is dedicated to Michael Large whose expertise and vision made the project possible. 
The MOST is operated by the School of Physics with the support of the Australian Research Council and the Science Foundation for Physics within the University of Sydney.
% WISE
This publication makes use of data products from the Wide-field Infrared Survey Explorer, 
which is a joint project of the University of California, Los Angeles, and 
the Jet Propulsion Laboratory/California Institute of Technology, 
funded by the National Aeronautics and Space Administration.
% 2MASS
This publication makes use of data products from the Two Micron All Sky Survey, which is a joint project of the University of 
Massachusetts and the Infrared Processing and Analysis Center/California Institute of Technology, funded by the National Aeronautics 
and Space Administration and the National Science Foundation.
%USNO
This research has made use of the USNOFS Image and Catalogue Archive
operated by the United States Naval Observatory, Flagstaff Station
(http://www.nofs.navy.mil/data/fchpix/).
% TOPCAT
TOPCAT\footnote{\underline{http://www.star.bris.ac.uk/$\sim$mbt/topcat/}} 
\citep{taylor05} for the preparation and manipulation of the tabular data and the images.

\end{document}